\documentstyle[12pt]{article}
\input epsf

\begin{document}\setlength{\unitlength}{1mm}

\def\question#1{{{\marginpar{\small \sc #1}}}}
\newcommand{\gsi}{\,\raisebox{-0.13cm}{$\stackrel{\textstyle>}
{\textstyle\sim}$}\,}
\newcommand{\lsi}{\,\raisebox{-0.13cm}{$\stackrel{\textstyle<}
{\textstyle\sim}$}\,}

\rightline{chao-dyn/9908012}
\rightline{July 1999}
\baselineskip=18pt
\begin{center}
{\large Stationary determinism in Observed Time Series: the earth's surface temperature}\\
\vspace*{0.2in}
{Rafael M. Gutierrez\footnote{Part of this work was performed as a Ph.D. dissertation at the Physics
Department at New York University in the Applied Science program.} \\
{\it Centro Internacional de F\'{\i}sica, CIF. \\ Edificio Manuel Ancizar, Ciudad Universitaria. \\ 
Santaf\'{e} de Bogota,Colombia}}\\

\vspace{.1in}
\end{center}
\vskip  0.1in  

{\footnotesize
 {\abstract
In this work we address the feasibility of estimating and isolating the 
stationary and
 deterministic content of observational time series, {\bf Ots}, which in general
 have very limited characteristics. In particular, we study the valuable 
earth's surface mean temperature time series, {\bf Tts}, by applying several 
treatments intended to isolate the stationary and deterministic content. We
 give particular attention to the sensitivity of results on the different 
parameters involved. The effects of such treatments were assessed by means 
of several methods designed to estimate the stationarity of time series.  
In order to strengthen the significance of the results obtained we have 
created a comparative framework with seven test time series of well-know 
origin and characteristics with a similar small number of data points. We 
have obtained a greater understanding of  the potential and limitations of 
the  different methods when applied to real world time series. The study of the 
stationarity and deterministic content of the {\bf Tts}  gives useful information 
about the particular complexity of global climatic evolution and the 
general important problem of  the isolation of  a real system from its
 surroundings by measuring and treating the obtained observations without
 any other additional information about the system.  } }

\section{ Introduction}
Correlations in time series have different origins. The origin of
 short term correlations of real systems with low-frequency divergences,
 called 1/f noise or colored noise, is unknown \cite{BP97b,Weiss88}. Artificial
 colored noise time series can be generated by introducing  in some manner
 short term correlations into an otherwise uncorrelated stochastic time 
series \cite{PSVM92}. Strong correlations at all time scales are typical of 
Brownian motion and can be artificially constructed by integrating a 
stochastic time series that has a Gaussian distribution. These different 
correlations are very common in nature and compete with the expected 
correlations generated by deterministic rules from the continuous processes 
underlying an Observational time series ({\bf Ots}). {\bf Ots} have in general several 
important limitations such as a small and sparse number of data points, 
poor resolution and accuracy, dynamical and measurement noise contamination.
 Dynamical noise affects each time step which may change the nature of the
 dynamics itself. Contamination with colored noise reinforces low 
frequencies in the power spectrum producing similar difficulties as those 
encountered when the time span of observation is too short to cover all 
the characteristic times of the system. When contamination comes from 
Brownian motion the time correlations extend to most of the data points of 
the time series and there is no time scale at which the static properties 
can be estimated. 
Therefore, the lack of stationarity in {\bf Ots} is a very common problem 
although, it is in general an implicit condition for the application of
 most of the methods used to 
identify and characterize determinism in time series. This implicit 
stationarity condition corresponds to the physical aspect of a defined 
time scale of observation within which all the relevant processes are 
followed and sufficiently sampled. 

The mentioned limitations of 
most {\bf Ots} combined with the actual complexity of the observed process and 
the strength of the interactions with other systems, frequently produce 
effects that can be seen as contaminations and lack of stationarity. These 
artifacts of the observation may be cleaned in some extent and hence, 
important dynamical information of the process may be isolated, 
characterized and eventually modeled. 

From a dynamical point of view 
nonstationarity may correspond to nonautonomous systems, changes of the 
evolution equations, changes of parameters with time, or even changes of 
the shape of the attractor effectively leaving unchanged the dynamical 
invariants. Therefore, we may be interested in the problem of extracting a 
meaningful stationary deterministic time series from the original {\bf Ots}. In
 order to achieve this goal we first must be able to estimate the 
stationarity of an {\bf Ots}. 
Since {\bf Ots} are irreproducible, the importance of stationarity in {\bf Ots} corresponds 
to the integrity and permanent identity of the system within the time it is observed 
with respect to the time resolution in which it is observed. Since we do not know a 
priori the deterministic rules and parameters characterizing the observed system, the 
{\bf Ots} is considered stationary if probabilities of transition between different states 
of the system are constant during the observed period of time \cite{KS97b}. From a 
practical point of view stationarity ensures the feasibility of a statistical analysis, 
since the probabilities are well defined as occurrence limits 
\cite{BP97b}. However,  linear statistics of chaotic time series such as the variance and mean 
do not approach finite values since they do not have renormalizable probability density 
distributions. Usually, {\bf Ots} have small a number of data points and the statistical analysis 
has in general additional limitations. 

An appropriate period of time and frequency of sampling may adequately manage the complex compromise 
between the system and the observational characteristics and we may consider the {\bf Ots}  stationary 
in practical terms. Unfortunately such particular time ``lens" is not easy if ever possible 
to determine, as several coexistent and nonseparable time scales may be involved. 

The earth's mean surface temperature time series, {\bf Tts}, is a particularly complicated case 
of {\bf Ots} in two ways. As a time series it is a highly contaminated and relatively small set 
of low resolution data points. From the point of view of the system that {\bf Tts} represents, 
the complexity of the underlying process may be excessive to be able to be characterized with the 
available data and the capability of the existing methods. However, the fact that each data point 
is a monthly earth's surface mean, considerably reduces the space and time scales concerned. 
Under such conditions many temporal and local complex processes of the earth's surface are 
conveniently rounded out. In this context it may be possible to identify certain global 
low dimensional dynamics where several complex processes make contributions without being direct 
deterministic aspects ( active degrees of freedom) of such global dynamics. 
This perspective permits to exploit some potential 
information about the earth's climatic evolution contained in this valuable time series obtained 
in more than a hundred years of measurement and data processing effort. Nevertheless, if there is not 
a particular set of scales and certain treatments permitting to obtain a time series where the 
``signal" dominates the different ``contaminations" the system is not observable. 

Stationarity is a property which can never be positively established in a finite time series. 
In practice there are some methods which give us partial measures of  how strong is the 
violation of stationarity of a given time series. Contamination and nonstationarity are not 
disconnected problems in particular for nonlinear dynamical systems. In general, for nonlinear 
time series it is not possible to filter out contaminations without affecting the actual dynamics 
because in some sense most of them are part of the dynamics itself. Consequently, we must try to 
measure how robust the dynamics is to external effects, how discernible these effects are from the 
actual signal and thus, estimate to which extent such dynamics can be extracted from the {\bf Ots}. For 
nonlinear systems when nothing substantial is known about the signal nor about the contamination, 
we need to characterize the signal and/or the contamination in some fashion which allows us to 
differentiate them by making certain assumptions. However, this goal has not been as successful 
as in the case of linear systems. There are only general criteria which can be used to distinguish 
which assumptions are better than others. For example, in {\bf Ots} of natural phenomena where in general 
colored noise is present, it is more efficient to make a local low pass filter than to assume white 
noise \cite{Abar97b}.

Most of the modern methods of nonlinear dynamics used to study time series involve a 
reconstruction of the scalar time series into $m$-dimensional data points in a Euclidean space. 
A norm is then used to measure distances and geometrical properties of such reconstruction. 
Embedding is a method to reconstruct the attractor of a system by transforming the scalar 
time series of observations 
${x_{i}}$, $i=1,..., N$, where $N$ is the total number of data points, into $m$-dimensional reconstructed 
vectors or points
 $\vec{x}_{i} = ( x_{i}, x_{i+\tau},..., x_{i+\tau(m-1)})$, where 
$i+\tau(m-1) \leq  N$ \cite{Abar97b} --\cite{ OSY94}. Such a reconstruction involves two 
important parameters, the time delay between each component of a reconstructed vector, $\tau$ , and the 
embedding dimension $m$. The demands on $\tau$ are not precisely defined by the embedding theorem. In 
practice, for {\bf Ots}, the choice of a correct $\tau$ may be very important because of the compromise between 
the relevant time scales of the system and the characteristics of the time series. When the attractor 
is correctly reconstructed any static geometrical characteristic of the system can be estimated. 
The embedding theorem is based on correlations from deterministic rules of the system that in principle 
exist in any macroscopic physical process. In the case of a chaotic system the trajectories remain confined 
to the attractor after temporal behaviors have died out. This attractor is a static structure, but when the 
temporal deterministic correlations are affected by parameter variations or changes in the deterministic rules 
of the dynamic or any other disturbing correlation, the stationarity of the system may be destroyed and the 
attractor is not static; it cannot be defined. The maximum norm takes the absolute value of the maximum 
component as the norm of the vector or $m$-dimensional point. The maximum norm is very economic in computational 
time but may involve certain difficulties when $m$ is small and the characteristics of the time series 
are limited as used to be the case for {\bf Ots}. This problem will be discussed within the application of 
some nonlinear methods to identify nonstationarity in a time series. 

The characteristics of the system itself and those of the corresponding {\bf Ots} determine how successful 
and how complicated could be the process of obtaining a meaningful stationary and deterministic time series 
from an {\bf Ots}. For nonlinear time series the success in such process has been very limited although, it is 
in constant development. In order to attempt such isolation for the {\bf Tts} we have chosen those methods with 
more general applicability or especially designed for nonlinear time series. Methods based on strong 
assumptions may generate artificial results when applied to {\bf Ots}. A particular limitation of {\bf Ots} is that 
we cannot manipulate certain parameters such as the sampling time and time span in order to improve 
stationarity without compromising the number of data points for statistical sufficiency and/or reducing 
the frequency of observations introducing contaminating effects.
 The methods used to obtain a meaningful stationary-deterministic time series from an {\bf Ots} can be classified 
in two groups. First, the methods applied in the time domain: \\
  \\
-Polynomial detrending which is natural to the time series and do not assume nonstationarity produced by low frequencies as is the case  for detrending with dominant frequencies. \\
-Low-frequency filter by introducing a decaying memory function in an integral-differential 
embedding \cite{Gilm98,Press92}. \\
-Smoothing with a different number of neighbors, or other moving average or FIR filters, 
which do not have a direct effect on stationarity but increase resolution and reduce 
fast variations \cite{Abar97b,CH92}--\cite{LR89}. \\
-First differences of data points removes slowly varying trends in the 
time series. \\
-The method of the correlation matrix filters out noise in the time domain in particular high 
frequency contaminations \cite{Gilm98,AHLS98}--\cite{RS92}. \\
-Noise reduction by nonlinear methods or local maps using neighborhood to neighborhood 
information \cite{KS97b,Abar97b,Sau92,KY90}--\cite{EBP93}. \\
  \\
Secondly, methods applied in the frequency domain: \\
   \\
-Windowing or low and high frequency filters are extensively used \cite{Press92}. \\
-Hilbert transform by using the FFT which is like the generalized zero derivative also 
used to interpolate small time series \cite{Gilm98}. \\
-Maximum entropy or all poles method, also used to detrend time series, estimates the 
power spectrum by representing the data in terms of a finite number of complex poles of 
discrete frequency instead of the essential polynomial fit used by the Fourier transform, 
especially useful when the data looks very noisy \cite{Laeri90}. \\
-Fourier interpolation permits to extend and clean out the data sets \cite{Gilm98,Press92}. \\

The most appropriate method or combination of methods which best isolates the stationary determinism 
potentially contained in a given {\bf Ots}, minimizing the damages of the dynamical information, is not easy 
to find and strongly depends on each particular case. For example, if the power spectrum is not small 
near the Nyquist frequency, frequency domain processing is a bad idea and should be abandoned \cite{Gilm98}. 
We may economize a lot of effort if we make the assumption that the outcome of the temperature measurements 
of the {\bf Tts}, or any other {\bf Ots}, is a random  variable implying a probabilistic approach which turns out to be 
very fruitful. However, it will not give as much information of the underlying mechanisms as a deterministic 
model may give. 

Given the strong difficulties to define a definitive method to measure and isolate stationary and 
determinism in {\bf Ots}, we have performed a comparative analysis of the {\bf Tts} and several treated {\bf Tts} altogether 
with some representative test time series of the same length and well known characteristics. In section 2 
we describe the {\bf Tts} and the test time series. In  section 3 we make a short presentation of several methods 
used to estimate stationarity in  finite time series placing particular attention on the problems 
generated by the limiting characteristics of the {\bf Ots}. In section 4 we present the results of applying 
several methods of filtering and detrending to the {\bf Tts}. We then, use a weak stationarity criterion in 
order to choose the most promising  treated {\bf Tts} ({\bf TTts}). In section 5 we apply the different methods 
presented in section 3 to the {\bf Tts}, the test time series and the chosen  {\bf TTts} obtained in section 4. 
In section 6 the results are summarized and discussed. In section 7 we present some conclusions.

\section{ The Temperature time series ({\bf Tts}) and test time series}
OAK Ridge National Laboratory provides synopsis of frequently used global change data \cite{Trends93}. 
Of this valuable information, we will use the global temperature anomalies obtained from instrumental 
surface air temperature records compiled by K. Ya. Vinnikov, P. Ya. Groisman and K. M. Lugina. 
This data has been mainly taken from the world Weather Records, Monthly Climatic Data for the World, 
and Meteorological Data for Individual Years over the North Hemisphere Excluding the USSR. This 
information was completed and improved using other data and methods that are explained and referred 
to in reference \cite{Trends93}. The original time series of 1356 data points corresponds to the 12 
monthly mean temperatures from 1881 to 1993, in chronological order. This data is given with two 
decimal digits of accuracy. For a more detailed discussion of the data and its limitations, noise 
and possible effects on the estimated averages the reader is referred to the original article of 
Jones, Wigley and Kelly \cite{JWK82}.

We are going to use seven test time series that will be identified by the names: {\bf Noise}, 
{\bf Random}, {\bf Pnoise}, 
{\bf Chaos}, {\bf Brownian}, {\bf Eeg} and {\bf Ekg}. These test time series, all with the same 1356 data points, have been 
obtained from CDA pro, a software for nonlinear analysis of time series \cite{CDA} where more details 
of the test time series can be consulted. {\bf Noise} is a time series of random  numbers with Gaussian or 
normal distribution centred on zero and a standard deviation of 1 called Gaussian or white noise. 
{\bf Random} is a time series of random  numbers uniformly distributed. {\bf Pnoise} is a time series of random  
numbers with Gaussian distribution centred on zero with 1/f power law. {\bf Chaos} is a time series of the $x$ 
variable of the Lorenz attractor. {\bf Brownian} is a Brownian time series generated by the integral of a 
Gaussian noise time series. {\bf Eeg} and {\bf Ekg} are two real world time series with expected nonstationarity. The first 
consists of measurements of the voltage in the electrodes of an electroencephalogram. The second 
consists of the intervals between successive interbeats of an electrocardiogram.

The {\bf Ots} are real world time series corresponding to measurements of a particular observable of  a 
natural system which in many cases are not repeatable and are irreproducible. They differ from 
experimental time series because they are not produced by an experimental set up which can be 
controlled to some extent, and from numerical time series because we do not know the equations 
which generate them. Many {\bf Ots}, in particular {\bf Tts}, correspond to systems that never will be in 
the same circumstances and therefore we have to limit ourselves to observe them as they evolve. 
We know that no real world system can be completely isolated from its surroundings; to some extent 
the environment and the observation process are part of the system itself. Similarly, no particular 
time scale can be completely isolated  from the others. Different parts of the system evolve at 
different frequencies with complex interrelations but conforming the same dynamics. In the case 
of the {\bf Tts} we have an external periodic forcing force of astronomical origin, that of the seasonal 
oscillation determined by the earth's orbit around the sun and the tilt of the rotation axis (precession). 
It is known that strong periodicities tend to hide underlying fractal structures \cite{BP97b,AY81} and 
important nonlinear and potentially chaotic dynamics may not be easily detected. Formally speaking, a 
periodic external force acting on the system may be included as a variable recovering a stationary point 
of view of the system. The importance of the periodicity observed in local seasonal variations of the 
temperature is evident. What is not so evident is how strong it is in the sense of global earth's surface 
mean temperatures. From the point of view of global means these seasonal cycles are indirectly generated 
by the asymmetry of those geological masses of the earth which permit the transformation of the incoming 
solar energy into thermal energy on the surface of the earth. The most important of such geological masses 
are earth, sea, ice, clouds and some factors related with the existence of life and particularly with human 
activities. These different masses and different activities strongly interact in complex ways.

\section{ Linear statistics and weak stationarity}
One reason for the actual incapacity to precisely measure the stationarity of a time series is that 
there is no uniform definition of stationarity. Loosely speaking, physical stationarity means the 
constancy of the main features of the system, dynamical stationarity means constancy of equations 
and parameters involved in an autonomous dynamical system, and statistical stationarity is defined 
for stochastic processes when joint probability functions are independent of time shifted 
realizations \cite{KS97b,Abar97b}. Nevertheless, these different perspectives of stationarity have 
close links. The deterministic aspects of a physical system are not conflictive with the contemplation 
of its signal as a realization of a stochastic process on which the statistical stationarity is based. 
In the language of dynamical systems statistical stationarity is equivalent to the existence of an 
invariant ergodic measure \cite{ER85,WKP98}.
If linear statistics like the mean, standard deviation and variance, estimated from different parts of a 
time series do not vary beyond statistical fluctuations, the time series may be considered weakly 
stationary. Weak stationarity does not imply dynamical stationarity but may be a useful first indication 
of it \cite{Ma98,GNG98}. Vice-versa, weak nonstationarity not necessarily implies dynamical 
nonstationarity, it may be produced for example, by a variety of artifacts in the processes of 
acquisition and manipulation of data. 

We cannot be certain of the effects of most of the detrending 
and cleaning treatments on the potential nonlinear dynamical information contained in a {\bf Ots}. However, 
we may consider weak stationarity as an initial criterion to assess the effects of different treatments 
applied to the {\bf Tts}. We use the variance V, as the linear statistics which is more sensitive than the mean, 
mean deviation and standard deviation but not too sensitive 
to statistical changes in the time series. The sensitivity of the linear statistics is important because 
of the limited number of data points but can become a problem if the time series is highly contaminated.
In order to test weak stationarity we divide each time series in three consecutive and nonoverlapping  
segments of equal number of data points. Comparing the first with the second segment of the {\bf Tts} we 
observe that V changes in ~53\%, and comparing the second with the third segment and the third with 
the first segment, the absolute changes of V are ~70\% and ~26\% respectively. These results represent 
an important weak  nonstationarity of the {\bf Tts}. The time series {\bf Chaos} present good weak stationarity with 
unimportant variations of V for the three segments of the data. The time series {\bf Noise} and {\bf Random}  present 
similar results. This is expected because these three time series are stationary, deterministic and 
stochastic respectively. The time series {\bf Pnoise} presents mild absolute variations of V as expected 
because of the short term correlations of this stochastic time series. 
The time series {\bf Eeg} and {\bf Ekg} give the expected weak nonstationarity 
more remarkable in the second case but, in both cases, it is considerably smaller than that 
of  {\bf Tts}. Table 1 summarizes these results. 

Within this context we have applied several methods to the {\bf Tts} in order to obtain Treated Temperature 
time series, {\bf TTts}, with improved weak stationarity. All of the {\bf TTts} were divided in three segments and 
analyzed as mentioned above. Those {\bf TTts} which present certain reduced absolute variations of  V compared 
with the corresponding values for {\bf Tts}, are considered {\bf TTts} with improved weak stationarity. We have 
considered six methods which do not need to make strong assumptions about the system and contamination 
of the time series. The first three methods mainly treat the noise contamination. The following two 
methods mainly treat the trends of the time series and, the last method removes
 slow trends and reduces some noise contamination. The six methods used (and their parameters) are:\\
  \\
   i) filtering with a correlation matrix (the size of the matrix), \\
  ii) smoothing (number of neighbors and variable intensity for different parts of the time series), \\
 iii) nonlinear filters ( $\tau$ , $m$ and number of neighbors),\\
  iv) detrending with the maximum-entropy or all poles method (number of poles),\\
   v) polynomial detrending (degree of the polynomial),\\ 
  vi) and first differences.\\ 

From an extensive list of more than seventy {\bf TTts} obtained by different combinations of the six methods 
with different values of the parameters involved, we have chosen the 21 {\bf TTts} with most important 
improvement of weak stationarity. These 21 {\bf TTts} are described below from 2 to 22 leaving 1 for 
the original {\bf Tts}. The results of weak stationarity are presented in table 1. \\
  \\
 {\bf TTts} 2 is {\bf Tts} smoothed with one neighbor considering a decreasing factor from 1 to 0.001 to compensate the assumption that the quality in obtaining and processing the {\bf Tts} has been improved with time.\\
 {\bf TTts} 3 is {\bf Tts} detrended with the maximum-entropy method using 128 poles. \\
 {\bf TTts} 4 is {\bf Tts} filtered with the correlation matrix where the three dominant modes have been used.\\
 {\bf TTts} 5 is {\bf TTts} 3 smoothed with the decreasing factor. \\
 {\bf TTts} 6 is {\bf TTts} 4 smoothed with the decreasing factor. \\
 {\bf TTts} 7 is {\bf TTts} 4 detrended with the method of the maximum-entropy using 128 poles.\\ 
 {\bf TTts} 8 is {\bf TTts} 4 detrended with the method of the maximum-entropy using 128 poles and finally smoothed with the decreasing factor.\\
 {\bf TTts} 9 is {\bf Tts} filtered with the nonlinear method in one dimensional embedding and using two neighbors. \\
 {\bf TTts} 10 is {\bf Tts} filtered with the nonlinear method in six dimensional embedding and using four neighbors. \\
 {\bf TTts} 11 is {\bf Tts} detrended with the method of the maximum-entropy using 128 poles and then filtered with the nonlinear method in one dimensional embedding and using two neighbors. \\
 {\bf TTts} 12 is {\bf Tts} detrended with the method of the maximum-entropy using 128 poles and then filtered with the nonlinear method in one dimensional embedding and using four neighbors. \\
 {\bf TTts} 13 is {\bf Tts} detrended with the method of the maximum-entropy using 128 poles and then filtered with the nonlinear method in six dimensional embedding and using four neighbors. \\
 {\bf TTts} 14 is {\bf Tts} detrended with a polynomial of order 2. \\
 {\bf TTts} 15 is {\bf Tts} detrended with a polynomial of order 3. \\
 {\bf TTts} 16 is {\bf Tts} smoothed once and then detrended with a polynomial of order 2.\\ 
 {\bf TTts} 17 is {\bf TTts} 4 detrended with a polynomial of order 2. \\
 {\bf TTts} 18 is the first differences of {\bf Tts}. \\
 {\bf TTts} 19 is {\bf Tts} with a nonlinear filter with embedding 1 and 1 neighbor. \\
 {\bf TTts} 20 is {\bf Tts} with a nonlinear filter with embedding 3 and 2 neighbors.\\ 
 {\bf TTts} 21 is {\bf TTts} 19 detrended with a polynomial of order 2. \\
 {\bf TTts} 22 is {\bf TTts} 20 detrended with a polynomial of order 2. \\

Some comments about the results of all the {\bf TTts} studied are in order to justify the 21 selected {\bf TTts}. 
The best results of weak stationarity using polynomial detrending are obtained for polynomials of order 
two and three, for higher order polynomials the weak stationarity degrades very fast. Smoothing degrades 
weak stationarity when we use more than one neighbor. This is an expected result because smoothing mainly 
cleans up homogeneously distributed contaminations which is not necessarily the case for nonstationary 
effects. In principle the first difference of the time series gives greater density in phase space and 
clarifies nonlinearities reducing the effects of any short-time linear correlation giving some times 
the same results than the original time series \cite{SM90}. For {\bf Tts} it clearly improves weak stationarity 
but other undesirable effects are produced by the low accurate and other limitations of such {\bf Ots}. 
Nonlinear filters have been conceived for potentially chaotic time series and they do not assume 
any particular knowledge about contamination. However, the scalar time series have to be reconstructed by 
an embedding procedure which demands information about the dynamics involved, in particular the embedding 
dimension which is not easy to estimate for {\bf Ots}. We have explored nonlinear filters of {\bf Tts} with embedding 
dimensions from 1 to 6. The {\bf TTts} obtained with a nonlinear filter present improved weak stationarity for 
small embeddings, in particular for dimension 1 and 3, and for small number of neighbors, 1 and 2. For 
larger values of these two parameters the results degrade quickly. {\bf TTts} obtained by combinations of 
different methods in general do not improve the results because of the low resolution and highly 
contaminated data points constituting the original {\bf Tts}. The difficulty to discriminate between signal 
and noise is exacerbated when the {\bf Tts} is manipulated repeatedly and round-off errors propagate 
indiscriminately. Therefore, the pretention to isolate the stationary deterministic content of a 
{\bf Ots} by applying successive treatments on the data may produce the opposite effect hiding most of 
such relevant and valuable information.

\begin{table}
\begin{center}
\begin{tabular}{|l|c|c|c|c|c|c|c|c|}            \hline
time series  & V 1-2 & V 2-3 &  V 3-1  & trend  &   period.  &  $t'$ &  freq.  &  sat. time    \\  \hline
1        &  2  &   3  &   4  &  5  &  6  &   7  &  8    &  9   \\  \hline \hline
Chaos    &  2  &   2  &   2  &  3  &  C  &   6  & .03   &  10*    \\
Noise    &  1  &   1  &   1  &  2  &  /  &   1  & -     &  1    \\
Random   &  3  &   5  &   1  &  1  &  /  &   1  & -     &  1    \\
Brownian & 13  &  38  &  42  &  26 &  C+ & 169  & 0     & non     \\
Pnoise   & 15  &  23  &  25  &  15 &  C- &  11  & 0     &  10    \\
Eeg      & 16  &  15  &   5  &  1  &  C  &   6  & .032   &  10    \\
Ekg      & 36  &  27  &  23  &  6  &  C  &   7  & .002  &  30    \\
Tts   1  & 53  &  70  &  26  &  13 &  C  &  77  & 0     &  12    \\
TTts  2  &  1  &  15  &  13  &  1  &  /  &   1  & -     &  1    \\
TTts  3  & 21  &  19  &  55  &  1  &  /  &   1  & -     &  1     \\
TTts  4  & 54  & 125  &   4  &  19 &  C  &  90  & 0     &  20      \\
TTts  5  & 10  &   2  &   8  &  7  &  /  &   1  & -     &  1    \\
TTts  6  & 12  &   5  &  14  &  5  &  /  &   1  & -     &  1    \\
TTts  7  & 23  &  21  &  64  &  1  &  /  &   1  & -     &  1    \\
TTts  8  &  3  &   9  &   5  &  1  &  /  &   1  & -     &  1    \\
TTts  9  & 31  &  28  &  14  &  12 &  C  &  14  & 0     &  20    \\
TTts 10  & 44  &  97  &  10  &  16 &  C  & 132  & 0     &  20    \\
TTts 11  &  4  &   5  &   2  &  1  &  /  &   1  & -     &  1    \\
TTts 12  &  4  &   8  &  11  &  1  &  /  &   1  & -     &  1    \\
TTts 13  &  2  &   8  &   9  &  1  &  /  &   1  & -     &  1    \\
TTts 14  & 56  &  53  &  53  &  1  &  C  &   6  & .01   &  20    \\
TTts 15  & 57  &  31  &  77  &  1  &  C  &   5  & .01   &  20     \\
TTts 16  & 58  & 102  &  17  &  1  &  C  &  10  & .01   &  20    \\
TTts 17  & 57  &  80  &  29  &  1  &  C  &   8  & .01   &  20   \\
TTts 18  & 56  &  37  & 228  &  1  &  C  &   1  & .25   &  1    \\
TTts 19  & 20  &  20  &   4  & 10  &  C  &  14  & 0     &  20    \\
TTts 20  & 40  &  62  &   2  &  32 &  C  &  76  & 0     &  20    \\
TTts 21  & 22  &  14  &  12  &  1  &  C- &   2  & .01   &  20    \\
TTts 22  & 45  &  49  &  21  &  1  &  C- &   4  & .007  &  20    \\  \hline
\end{tabular}
\caption{ {\footnotesize Column 1: name of the time series. Columns 2 to 4: absolute percentage change of 
the variance V, for the three pairs of 
consecutive equal segments of the corresponding time series respectively. Column 5: the maximum 
difference of the second order polynomial trend with the average value of the corresponding time series;
the value is given as a percentage of the range of values of the time series. Column 6: 
characteristic shape of the periodogram: ``/" indicates a flat power 
spectrum with periodogram along the 45-degree diagonal characteristic of an uncorrelated stochastic time series, 
and ``C" indicates curved or far from 
the 45-degree diagonal periodogram representative of a fast decreasing power spectrum. ``C+" and ``C-" 
are used to indicate that the periodogram is further or closer to the 45-degree diagonal which are 
characteristics of colored noise and Brownian time series respectively. Column 7: the 
correlation time $t'$ defined by $C(t')\sim C(0)/e$. 
Column 8: the dominant frequency of the power spectrum estimated by the method of minimum entropy 
using 128 poles; the Nyquist frequency, 0.5, is the maximum possible value and ``-" indicates that no 
dominant frequency is observed. Column 9: saturation time of  the STSP; * 
for the time series {\bf Chaos} the first maximum is considered the saturation time.}}

\end{center}
\end{table}

\section{ Methods to estimate stationarity in finite time series}
In accordance with the difficulties presented above we cannot expect a simple method to diagnose 
the stationarity of a given {\bf Ots}. Practical and 
conceptual problems 
are all related  with the compromises between the relevant scales of the observed 
system and the 
process of observation. Together with linear statistics, used to estimate weak stationarity, we 
apply some of the most important methods to identify and estimate the stationarity of a finite 
time series. We succintely describe these methods giving particular attention to the problems 
encountered when we applied them to {\bf Ots}:

1- \underline{Polynomial Fit}: The most natural method to identify nonstationarity 
is the direct inspection of the time series, 
we may identify the presence of trends and bursts. Bursts are very short time and large amplitude changes. 
Such bursts 
may correspond to faster dynamics  than the time resolution of the present time series and 
are only captured occasionally as eventual coincidences. Alternatively, the time series under study may 
have some dynamical information that can only be captured with a longer period of observation 
where a sufficiently large number of such bursts are included. An evident trend on the 
plot of the time series may correspond to slow variations that cannot be resolved by the 
time spanned by the {\bf Ots}. Trends of a time series can be quantitatively estimated by a polynomial 
fit and then the time series can be detrended by subtracting such fit. For an {\bf Ots} we do 
not know the appropriate order of the polynomial to be used and we are not certain of the effects 
on the nonlinear information of the time series.

2- \underline{The correlation function} or autocorrelation function $C(t)$, is a measure of 
time correlations 
between the different data points of a time series. $C(t)$ is a linear measure of temporal 
correlation in the sense of least squares prediction of data points at time $t$ from the
 knowledge 
of  data points at a previous time \cite{BP97b,Abar97b}. The correlation time $t'$, is commonly 
considered the time at which $C(t)$ goes to $1/3$ or $1/e$ of its value at $t=0$. When $t'$ is larger 
or on the order of the duration of the time series, it indicates certain nonstationarity. The linear 
correlation measured by $t'$ is not necessarily the most relevant correlation in a nonlinear 
dynamical system \cite{Papo84}. Uncorrelated stochastic time series
 present the expected $t' \sim 0$. 
The short term correlations introduced into an otherwise stochastic uncorrelated time series to 
produce artificial colored noise time series, is detected by $C(t)$ 
giving a non negligible $t'$. 
The continuity and derivativity of dynamical systems of differential equations generate time series 
with strong correlations which correspond to large values of $t'$. These characteristic results 
strongly depend  on  the sampling time or frequency of measurements. Chaotic time series obtained 
from nonlinear systems of differential equations show in general relatively large $t'$ of the order 
of the inverse of the Lyapunov exponent. However, the corresponding $C(t)$ decreases very fast 
consistently with the divergent nature of chaotic behavior. The value of $t'$ for  {\bf Ots} depends on 
several aspects concerning the system and the corresponding time series. In some cases $C(t)$ also 
permits to identify certain nonstationarities associated with intermittency if such intermittency is 
captured by time correlations of temporal neighbors in the form of striking recurrent revivals while 
decaying on the average.

3- \underline{The power spectrum} (PS), is closely related with $C(t)$. The information provided by these two 
methods may be very similar in particular for long-time regimes \cite{AS81}. In the case of {\bf Ots}, 
the long-time regime is rarely achieved and each method may give different information about certain 
stationary aspects of the underlying system. The standard power spectrum defined by the squared Fourier 
transform of the time series can also be defined in terms of $C(t)$ and both definitions coincide under 
mild restrictions given by the Wiener-Khinchin theorem \cite{BP97b}. This coincidence permits to relate 
fast decreasing behavior in a wide but finite frequency range of PS with slowly decaying $C(t)$ which is 
characteristic of long-range memory. For this decreasing range, Chaotic time series sometimes present 
exponential behavior while stochastic time series present power law behavior \cite{LD96}. For finite 
time series this distinction is in general very difficult, in particular for certain colored noise 
time series because the power spectra is not sensitive to the phase information of the Fourier transform 
and cannot provide a full characterization of the dynamics. There is not a satisfactory explanation for 
this behavior of the power spectrum of colored noise \cite{BP97b}.
The sampling theorem due to Nyquist \cite{Nyq28} and Shanon \cite{Sha49}, gives the conditions under which 
a continuous signal is completely determined by the discrete sampled values constituting the time series. 
The power of frequencies higher than the Nyquist frequency defined as the reciprocal of twice the sampling 
time must be zero, otherwise the problem of "alising" will be present \cite{EMc97b}. To avoid alising we 
may use a low-pass filter removing high frequencies or, alternatively, increase the sampling frequency. 
The sampling frequency may be augmented by means of an interpolation method in the time domain or fast 
Fourier transform-based interpolation \cite{Gilm98, Press92}. These methods are not strongly recommended 
for potentially nonlinear time series. Since PS reflects the contribution of all periods from the sampling 
interval up to the total time covered by the time series and the {\bf Ots} are commonly highly contaminated, it 
is normal to have the Nyquist and higher frequencies small but different from zero. This may correspond to 
a contamination amplitude from different noise sources and not necessary to under-sampling. 
Random (infinite) time series such as experimental error and numerical round-off, have an absolutely 
continuous PS \cite{BP97b}. On the other hand, colored noise has most of its power concentrated in low 
frequencies. Nonstationarity is commonly present when a considerable part of the spectrum of a time 
series is in low frequencies implying that most of the variations of the system have very few oscillations 
during the observation time, i.e. many relevant times scales of the dynamics are of the order of the total 
observation time. The low frequencies of the PS can also be affected by low resolution and slow forcing forces. 
If this behavior is persistent down to zero frequency, the system will present a singularity of the power 
spectrum. This nonintegrable PS, whose integral is proportional to the average energy of the system, implies 
a nonstationary system because it is obtaining energy from the outside. However, dissipation may provide the 
mechanism to generate a stationary low dimensional attractor in state space. The high frequencies of the PS 
can also be affected by white noise, short time effects, fast forcing forces and sub-harmonics. Thus, to 
improve the stationary dynamical content of a {\bf Ots}, we should in principle, filter both extremes of the 
power spectrum reducing all possible contaminations and nonstationarity. It is important to remember that 
for chaotic time series most of the interesting frequencies are  low frequencies around the characteristic 
cycle times and filtering out frequencies a factor of ten greater and a factor of ten smaller than the 
frequency corresponding to such cycle will work at least qualitatively well most of the time. Such 
cycle times are unknown in the case of a {\bf Ots} and are difficult to identify before improving the stationary 
deterministic content of the time series.

4- \underline{Nonlinear Statistics and complexity measures}: The convergence of nonlinear statistics is 
not very well known apart from its very slow convergence making it very difficult to obtain meaningful 
results for small {\bf Ots}. The correlation dimension $D_{2}$ \cite{KS97b,Abar97b,GP83,GP84}, is particularly 
sensitive to nonstationarity from parameter drift and in particular to insufficient sampling. This 
sensitivity could be useful to estimate the stationarity of a given time series. However, estimates of 
$D_{2}$ are also very sensitive to contamination, accuracy and size of the time series. Estimates of $D_{2}$ for 
sections of  small {\bf Ots} are therefore very inaccurate and no useful information for stationarity may be 
obtained from them. The capacity dimension $D_{1}$ \cite{KS97b,Abar97b,Mand82,FOY83}, is less sensitive 
to the limitations of {\bf Ots} than $D_{2}$. On the other hand, it is not as good an estimate of the attractor 
dimension. This is because $D_{1}$ only captures geometrical information while, $D_{2}$ captures geometrical 
as well dynamical information of the attractor. Initially we have used five different nonlinear 
statistics designed to identify determinism in apparently stochastic time series. We applied them 
in each case to the whole time series, and to only the first and second half of the data points to 
obtain a better statistical meaning of the results. The five nonlinear statistics are: \\
  \\
i) capacity dimension, $D_{1}$, \\
ii) largest Lyapunov exponent, $Le$ \cite{WSSV85,EKRC86}, \\
iii) algorithmic complexity measured by the Lempel-Ziv method, $LZ$ \cite{LZ76,KS87}, \\
iv) and $BDS$ statistics which measures deviation from pure randomness \cite{BS88}, \\
v) Hurst exponent, $He$, which is a measure of self similarity \cite{Feder88}--\cite{LY97}. \\
\\
5- \underline{Recurrence Density Plots} (RDP): Recurrence plots were introduced to identify nonstationarity 
in a time series when the underlying  dynamical system is not autonomous (time appears explicitly 
in the evolution equations), and when the characteristics times of such systems are of the order of the 
length of the time series \cite{EKR87}. Later works have extended the utility and applicability of such 
plots \cite{TGZW96}--\cite{Casdagli97}. In particular there is strong evidence, although not yet formally 
proved, that most of the qualitative as well as quantitative results do not depend on the embedding 
parameters at least for embedding dimensions smaller or equal to 3 \cite{IB98, Nota1}. This is a very 
interesting result because good estimates for the embedding parameters are not easily determined from 
{\bf Ots}. We have verified this hypothesis with the time series {\bf Chaos} with embedding dimensions from 1 to 5, 
and with the {\bf Tts} for embedding dimension from 1 to 4 and time delays 1 and 9 sampling times where 9 
corresponds to the first minimum of the average mutual information \cite{Abar97b, ABST93}. Only small 
numerical variations have been observed.  The density of distances, trend or recurrence density plot 
RDP \cite{EKR87, TGZW96}, is a kind of recurrence plot which measures the number of pairs of points 
within a threshold distance ($\sim 10\%$ of the range of the time series), as a function of their temporal 
separation. The RDP is particularly useful for {\bf Ots} because their characteristic small number and low 
accurate data points make it very difficult to observe patterns and varying densities in a traditional 
two dimensional recurrence plot. For an autonomous system with short characteristic times compared with 
the length of the time series, RDP must look globally constant. This global behavior may be superimposed 
to a smaller time scale structure which would correspond to cycles of the system. These cycles must be 
much smaller than the length of the time series in order to indicate stationarity. For a nonstationary 
time series RDP would present an overall decreasing behavior.

6- \underline{Cross Prediction Error} (CPE): The nonlinear cross prediction method was introduced 
by Schreiber in 1997 \cite{Sch97}. This method has a very important conceptual contribution because it 
is based on the similarity between sections of the time series itself rather than similarity of certain 
statistics estimates from different sections of the time series. This method may be qualitatively 
independent of the particular statistics used. This method is particularly useful when the nonstationarity 
emerges from changes of the shape of the attractor while dynamical invariants remain effectively 
unchanged \cite{KS97b,Sch97}. This method has the practical advantage of a small demand of data 
and more robustness to contamination. This is because it compares the data with itself using all 
of the data points and may use nonlinear predictions which demand rather short segments of data 
points. The nonlinear cross prediction method generates a plot of 
nonlinear Cross Prediction Errors (CPE), with two indices indicating 
the predicted and predictee segments of the time series used in each estimate. In this context the 
predicted segment is used to construct $m$-dimensional vectors and the predictee segment is used to 
construct the neighborhood for a locally constant approximation \cite{Sch97}. The whole time series 
is divided into equal, consecutive and nonoverlapping segments of length $l$. 
$l$ must be larger than the characteristic times of the process where all 
the typical cycles of the system are completed several times. 
For stationary time series and an appropriate length for $l$, we expect that CPE is independent of 
the indices. A purely 
random time series must present large and homogeneous CPE for all combinations of segments. Time 
series with correlations from deterministic, or any other origin, may present certain patterns in 
the CPE which depend on $l$ and indicate the time length of such correlations.
This method has the disadvantage of having four parameters and the results of the method may be very sensitive 
for some ranges of their values when applied to {\bf Ots}. These four parameters are the size of the segments 
$l$, in which the time series is divided, the distance to determine the neighbors of a given data point, 
$\epsilon$, and the two 
reconstruction parameters, the time lag $\tau$, and the embedding dimension $m$. The sparseness and 
resolution of the data are very important for the first two parameters. 
The range of CPE evidently depends on the range and resolution of the data points. The compromise 
between $\epsilon$ and the density of data points is important because it determines the number of 
neighbors from which the prediction is going to be made. If $\epsilon$ is small and the time series 
is dense the situation is ideal but, if the data is sparse we may not find close neighbors. If $\epsilon$ 
and $l$ are large and the data is dense we may be using neighbors from dynamically different parts of the 
system and confusing results can be obtained. {\bf Ots} are in general sparse time series therefore, $\epsilon$
cannot be small and $l$ cannot be large and we cannot easily reduce the risk of making predictions with 
points from dynamically different periods of the time series. If  $\epsilon$ is too large the predictee 
interval dominates the results of CPE and no information is obtained from the predicted interval. When 
the studied time series is from a known system these problems can be avoided by considering the 
characteristic cycles involved. 

7- \underline{The Space Time Separation Plot} (STSP), was introduced by Provenzale et al  \cite{PSVM92}. 
It can be used to estimate the correlation time of a scalar time series. This is made by 
comparing geometrical and temporal correlations giving information about the stationarity 
of the time series. STSP is robust to  most of the limiting characteristics of {\bf Ots} and has been 
used to obtain good geometrical estimates such as the correlation dimension. STSP measures 
the effects of time correlations on the probability of finding two points within a given distance. 
In practice, STSP estimates the correlation times existent in a time series by measuring the 
spatial distance to be covered in order to find certain predetermined proportion of points for 
different time separations \cite{KS97b}. If such spatial distance does not saturate for increasing 
time separation, all the data points of the time series are temporally correlated and no 
stationarity can be obtained for any time scale. If saturation is obtained for a certain time 
separation the time series can be considered stationary for time scales larger than this time 
separation. Despite the general robustness of STSP with respect to the conditions imposed by {\bf Ots}, 
the method is very sensitive to the percentage of points and the size of the spatial partition. 
These two parameters give the resolution of the plot to resolve the actual behavior from numerical 
variations. As a scatter plot of the spatial separation versus the time separation between points, 
STSP must not depend strongly on the embedding dimension $m$ \cite{IB98, Nota1}. It was first 
verified with the results for the time series {\bf Chaos}  where only small numerical variations were 
observed as $m$ was varied. The cycles of the Lorenz attractor around the two fixed points are 
seen as certain periodicity of the STSP on the order of $\sim 14$ sampling times, see figure 5a. For {\bf Ots} with low 
sampling rates we can eventually find near returns closer to a given point than its dynamical next 
neighbor. In this situation the nonstationarity of the time series given by long term correlations 
cannot be detected by the STSP, i.e. the plots saturate fast and remain constant for any value of 
time. For {\bf Ots} with low resolution data points it is still easier to find return points and even in 
the case of short {\bf Ots} the STSP may not be able to detect nonstationarity for long correlation times. 

We now mention some other important methods that are not going to be used in this work:  

i) Comparison of different linear or nonlinear  tests for several consecutive and overlapping segments 
of the time series increases the number of data points but the independence of each segment is reduced 
\cite{WKP98, IK93, MS96}. These strategies may not give useful results when applied to {\bf Ots} with a small 
number of data points. 

ii) The concepts introduced by the nonlinear cross prediction method can be applied in different forms  and with different statistics. 
For example, the identification of neighbors to make the nonlinear prediction can be weighted with 
different criteria \cite{SM90,LGKF93h}. However, these criteria are not general and may work for 
particular situations which must be identified in advance depending on the knowledge of the system 
which is not available for most {\bf Ots}.

iii) The surrogate data method to evaluate the results of nonlinear analysis of time series has been 
tested against non-stationarity, in particular for cyclostationarity \cite{Timmer99pp}.  Surrogate data 
is a robust method of general applicability that could be exploited in order to test more directly the 
stationarity of {\bf Ots}.

iv) An important method to test stationarity in time series that has not been applied in this work uses 
the information in the time distribution of points in a state space reconstruction \cite{Ken96}. The 
need of accurate estimates of the embedding parameters makes this method potentially problematic for {\bf Ots}. 
Another related method based on time domain information in reconstructed phase space \cite{YLH98}, presents 
no qualitative dependence on the embedding parameters and seems to work well for small time series. 
These are two important desirable qualities for a method used to estimate stationarity of {\bf Ots}. 
These methods
have to be included in further analysis to understand {\bf Ots} and in particular {\bf Tts}. However, it is 
important to measure their sensitivity to contamination, data resolution and other
important limitations of {\bf Ots}. 

v) Two novel methods to test stationarity in time series may give relevant results when applied to {\bf Ots}. 
They involve wavelets multi-scale ideas which are natural tools to study stationarity \cite{YSM99pp} and, 
powerful and well-studied data compression techniques which are robust to noise \cite{KM99pp}. Wavelet 
transforms can also be used to remove nonstationarity from time series that were affected at certain 
specific time scales or ranges of time scales by some particular circumstance \cite{AGH88}--\cite{TFT97h}.    

\section{ Results} 
The results obtained with the test time series give validity to the algorithms used for each method. 
These results also contribute to define a practical framework for the interpretation of the sensitive results 
corresponding to {\bf Tts} and the different {\bf TTts}.

\underline{Polynomial fit}: The value of the linear and quadratic constants in a  polynomial fit give a quantitative measure of the trend 
of the analyzed time series. If these values are large compared to the range of the time series, the linear 
and quadratic corrections are important compared with a detrended behavior. The stationary stochastic time 
series {\bf Noise} and {\bf Random}  present the expected small values and no evident trend is observed. The real world  
time series {\bf Eeg} and {\bf Ekg} do not present clear trends, only in the second case a small trend is observed. The 
stationary chaotic time series {\bf Chaos} presents a negligible trend probably caused by the small size of the 
time series. The nonstationarity of the time series {\bf Pnoise} and {\bf Brownian} is evident with large linear and 
quadratic constants. No trends are observed for the {\bf TTts} 2,3,7,8,11,12,13,14,15,16,17,18,21 and 22. Evident 
trends exist for {\bf Tts} and {\bf TTts} 4,9,10,19 and 20 which were not detrended by any method, only filtered. 
{\bf TTts} 5 and 6 present less evident trends. In table 1 column 5 we give a quantitative measure of the trends 
for all of the studied time series consisting on the maximum difference between the average value of the 
time series with the polynomial fit. This value is given as a percentage of the total range of the data 
of the corresponding time series. Apart from the time series {\bf Brownian} and {\bf Pnoise} which present very large 
constants in the quadratic term of the polynomial fit, all other time series studied present almost 
linear trends when nonnegligeable.
 
\underline{Correlation function}: In table 1 column 6,  we present the results of the correlation time $t'$ in 
units of the sampling time, for the seven test time series, the {\bf Tts} and the 21 {\bf TTts}. The continuous 
deterministic origin of the time series {\bf Chaos} generates a correlation, $t'$=5.99, which is 
small enough to be consistent with the chaotic nature of the time series. The real world time series 
{\bf Eeg} and {\bf Ekg} present similar values $t'$=5.62 and 7.12, respectively. The correlation time of the 
stochastic but temporally correlated time series {\bf Pnoise} is of the same order of magnitude with a slightly 
larger value, $t'$=11.33. The time series {\bf Brownian} presents a very large correlation time, $t'$=169.37, 
consistent with the construction of such stochastic time series. The stochastic and  
uncorrelated time series {\bf Random}  and {\bf Noise} present the expected result $t' \sim 1$ and $C(t)\sim 0$ for 
all $t>1$. {\bf Tts} presents a large correlation time $t'=76.58$. The low resolution of the data points of {\bf Tts} 
may produce apparently long term correlations as measured by $C(t)$. The {\bf TTts} 2,3,5,6,7,8,11,12,13 and 
18 have all stochastic uncorrelated-like behavior with $t'\sim 1$ and $C(t)\sim 0$ for all $t>1$. The treatments 
to obtain {\bf TTts} 4 and 10 increase the value of $t'$ obtained for {\bf Tts} and that of {\bf TTts} 20 leaves it 
unchanged.  The rest of the {\bf TTts}: 9,14,15,16,17,19,21 and 22 present $t'$ values close to those of 
the time series {\bf Chaos}, {\bf Pnoise}, {\bf Eeg} and {\bf Ekg}.

\underline{Spectral analysis}: The five test time series {\bf Chaos}, {\bf Pnoise}, {\bf Brownian}, 
{\bf Eeg} and {\bf Ekg}, and the {\bf Tts} present very 
small or zero dominant frequency depending weakly on the number of poles used in the minimum entropy method. 
The time series {\bf Noise} and {\bf Random}  do not present a dominant frequency. More quantitative information is 
obtained from the cumulative periodogram which is the integral of the power spectrum over frequency. A 
cumulative periodogram following the 45-degree line, as is the case for the two uncorrelated stochastic 
time series {\bf Noise} and {\bf Random}, indicates the flatness of the spectrum. The other five test time series 
present far from 45-degree line periodograms except for the temporally correlated time series {\bf Pnoise} 
which is clearly closer but not along this line. The periodogram for {\bf Tts} is far from the 45-degree line 
being an important qualitatively difference with the time series {\bf Pnoise}. 
The {\bf TTts} 2,3,5,6,7,8,11,12 and 13 present a flat power spectrum verified by the cumulative periodogram 
following the 45-degree line characteristic of uncorrelated stochastic time series. The procedures 
generating the {\bf TTts} 4,10,19 and 20 do not produce any identifiable difference of the power spectrum 
compared with that of the {\bf Tts}. The power spectrum of the remaining {\bf TTts} 9,14,15,16,17, 21 and 22 present 
small changes compared to that of the {\bf Tts} with the exception of {\bf Tts} 18 where the periodogram is inverted. 
In table 1 column 6 we summarize these results qualitatively by using ``C" for curved or far from the 
45-degree line periodograms and ``/" for periodograms along this line. When the periodogram of a {\bf TTts} 
gets closer to the 45-degree line than the periodogram of the {\bf Tts}, this may indicate that the corresponding 
treatment generates some losses of the potential dynamical information contained in {\bf Tts}. The case of {\bf Tts} 
18 presents a strongly damped power spectrum in low frequencies with most of the power concentrated in high 
frequencies. This increases the possibilities of stationarity but also indicates an increase of contamination 
probably introduced by numerical round-off and the low accuracy of {\bf Tts}.
The dominant frequency remains zero only for the {\bf TTts} 4,9,10, 19 and 20. The {\bf TTts} 14,15,16,17,21 and 22 have 
a dominant frequency $ \sim 1/50$ of the Nyquist frequency.  Except for {\bf TTts} 18, the remaining {\bf TTts} present 
no dominant frequency and a flat PS. The dominant frequencies are presented in table 1 column 8.
 The linear decay in a log-linear scale of the power spectrum corresponds to exponential decay in linear-linear 
spectrum frequently associated with a chaotic time series which is clearly observed for the time series {\bf Chaos}. 
This characteristic is hard to precise objectively for {\bf Ots} but certain tendency can be identified. The 
{\bf TTts} 4,14,17, 21 and 22 present an improved linear decay compared with that of the {\bf Tts}. For the test 
time series the two extremes are clear, linear decay for the time series {\bf Chaos} and a flat PS for the 
uncorrelated stochastic time series {\bf Noise} and {\bf Random}. For the correlated stochastic 
time series {\bf Pnoise}, 
{\bf Tts} and the real world {\bf Eeg} and {\bf Ekg} the behavior of PS is between these two 
well defined behaviors with a 
certain power law or 1/f spectrum.    
The problem of alising and therefore, the applicability of the sampling theorem, has been apparently solved 
for all of the studied time series except for {\bf TTts} 18. As far as stationarity is concerned we are more 
interested in the low frequency part of the power spectrum where we can observe indications of the problems 
of low sampling and slow trends. 

\underline{Nonlinear statistics and complexity measures}: As anticipated, the small size of the time series studied in 
this work do not permit to obtain significant results for stationary analysis using nonlinear statistics 
on sections of such small time series. Only for the Lempel-Ziv complexity measure ($LZ$), Hurst exponent ($He$),
 and the BDS statistics ($BDS$), we have obtained certain results that 
deserve to be mentioned. The seven test time series present the expected results with acceptable accuracy 
giving some validity to the results obtained for the {\bf Tts} and the {\bf TTts}.  These results are presented in 
table 2 for the whole time series and the corresponding first and second halves indicated by I and II 
respectively.
The uncorrelated stochastic time series {\bf Noise} and {\bf Random}  have $LZ\sim 1$, $He\sim 0$ and $BDS <<0$. The correlated 
time series {\bf Pnoise} has $LZ\sim 0.8$, $He\sim 0.16$ and $BDS\sim -0.5$ which indicates a less complex 
time series, more 
self-similar and less stochastic than the two previous time series. The long term correlations of the time 
series {\bf Brownian} reduce the complexity of the otherwise uncorrelated stochastic time series to $LZ\sim 0.2$, 
increases self-similarity to $He\sim 0.5$ and destroys its stochastic or random  nature to $BDS\sim 2.0$. The time 
series {\bf Chaos} is much less complex $LZ\sim 0.2$, strongly self-similar $He\sim 0.5$ and 
clearly nonrandom $BDS\sim 0.6>0$. 
The two real world time series {\bf Eeg} and {\bf Ekg} give similar results with mild complexity $LZ\sim 0.5$, relatively 
strong self-similarity $He ~< 0.4$ and clearly nonrandom $BDS \gg 0$. The different results for the whole time 
series and the corresponding two halves are consistent within (large) statistical variations. However, the small 
number of data points used in all cases, in particular when half time series are considered, do not permit 
to resolve the sensitivity of these statistics to stationary changes from the statistical variations.
The results for {\bf TTts} 2, 3, 5,6,7,8,11,12,13 and 18, and less clearly for {\bf TTts} 9,19,20,21 and 22, present 
very close results to those characteristics of stochastic time series. 
That is $LZ\sim 1$, $He\sim 0$ and $BDS \ll 0$ 
for the three values corresponding to the whole and two halves of each {\bf TTts}. The {\bf Tts} and  {\bf TTts} 4, 10,14,15,16 and 17 definitively 
survived the stochastic classification from the nonlinear statistics point of view.

\begin{table}
\begin{center}
\begin{tabular}{|l|r|r|r|r|r|r|r|r|r|} \hline
{\em time series} & \multicolumn{3}{c|}{ $LZ$} & \multicolumn{3}{c|}{ $He$}  & \multicolumn{3}{c|}{ $BDS$}  \\ \hline
 name  &  whole  &  I  &   II  &   whole  &  I  &   II  &  whole    &  I  &  II \\  \hline
Chaos    &  .19  &   .20  &   .24  &  .46  &   .50  & .50   &  .66 & .45 & .75    \\
Noise    &  1.06  &   1.09  &   1.05  &  -4e-3  &   -9e-3  & -4e-3  &  -5.9 & -4.1 & -4.2    \\
Random   &  1.02  &   1.0  &   1.01  &  -3e-3  &   -5e-3  & -4e-3   &  -16 & -10 & -9    \\
Brownian & .18  &  .25  &  .14  &  .53 & .53  & .51   & 2.0 & 1.9 & 1.3     \\
Pnoise   & .77  &  .81  &  .79  &  .16 &  .18  & .17   &  -0.5 & -1.0 & -0.7    \\
Eeg      &   .49  &  .5    &  .5    &  .44  &   .40  & .39   &  4.2 & 2.9 & 3.0    \\
Ekg      &  .51   &  .55    &  .54    &  .33  &   .33  & .36  &  3.6 & 2.5 & 2.7    \\
Tts   1  & .63  &  .75  &  .78  &  .13  &  .13  & .17   &  1.97 & -0.78 & 1.01    \\
TTts  2  &  1.03  &  1.07  &  1.06  &  7e-3  &   2e-3  & 3e-3   &  -5.6 & -6.8 & -7.0    \\
TTts  3  & 1.06  &  1.07  &  1.05  &  3e-3  &   1e-3  & 2e-3     &  -5.3 & -5.3 & -5.3     \\
TTts  4  & .44  & .63  &   .45  &  .27  &  .29  & .50     &  3.3 & 1.7 & 2.3      \\
TTts  5  & 1.06  &   1.08  &   1.07  &  3e-3  &   7e-3  & 6e-3   &  -10 & -7 & -8    \\
TTts  6  & 1.04  &   1.03  &  1.03  &  1e-2  &   1e-3  & 1e-3     &  -6 & -5 & -5    \\
TTts  7  & 1.08  &  1.03  &  1.04  &  1e-3  &   1e-3  & 1e-3     &  -10 & -9 & -8    \\
TTts  8  &  1.06  &   1.05  &   1.04  &  2e-3  &   2e-3  & 2e-3     &  -6 & -7 & -7   \\
TTts  9  & .81  &  0.99  &  1.01  &  4e-2  &  4e-2  & 6e-2     &  -2 & -3 & -2    \\
TTts 10  & .5  &  .7  &  .8  &  .15  & .15  & .18     &  2.8 & 3e-3 & 1.4   \\
TTts 11  &  1.02  &   1.03  &   1.03  &  -7e-3  &   -6e-3  & -5e-3     & -9 & -8 & -8    \\
TTts 12  &  1.06  &   1.05  &  1.04  &  -5e-3  &   -4e-3  & -4e-3     &  -8 & -7 & -7    \\
TTts 13  &  1.07  &   1.06  &   1.06  &  1e-3  &   1e-3  & 1e-3     &  -9 & -7 & -7    \\
TTts 14  & .84  &  .93  &  .74  &  .12  &   .13  & .17   &  -0.5 & -1.9 & 0.6    \\
TTts 15  & .81  &  .93  &  .71  &  .13  &   .13  & .17   &  -.06 & -1.8 & 0.2     \\
TTts 16  & .55  & .65  &  .49  &  .32  &  .35  & .40   &  2.8 & 2.0 & 2.5    \\
TTts 17  & .62  &  .64  &  .64  &  .27  &   .29  & .33   &  2.6 & 1.3 & 2.4   \\
TTts 18  & 1.03  &  1.02  & 1.01  &  -4e-2  &   -2e-2  & -3e-2   &  -8 & -8 & -7    \\
TTts 19  & .79  &  .94  &   .95  &  5e-2  &  5e-2  & 8e-2     &  -2.1 & -3.0 & -2.2    \\
TTts 20  & .71  &  .82  &   .91  &  .1  &  .1  & .1     &  1.1 & -1.8 & -0.4    \\
TTts 21  & .95  &  .99  &  .90  &  5e-2 &   5e-2  & 8e-2   &  -4 & -3 & -2    \\
TTts 22  & .86  &  .95  & .81  &  .1 &   .1 & .1  &  -2 & -3 & -1    \\  \hline
\end{tabular}
\caption{{\footnotesize Nonlinear statistics and complexity measures: Algorithmic complexity of Lempel and Ziv 
($LZ$), Hurst exponent as a measure of self-similarity ($He$), and the $BDS$ statistic as a discriminator from 
pure stochastic time series ($BDS$). The value corresponding to the whole time series is placed in the first 
column in front of the name of the corresponding time series and under the corresponding nonlinear 
statistics. The values corresponding to the first and second half of the same time series are 
placed in the following two columns indicated by I and II respectively, 
under the corresponding nonlinear statistics.}}
\end{center}
\end{table}

\underline{Recurrence Density Plots} (RDP): The test time series {\bf Chaos}, {\bf Pnoise}, {\bf Noise}, {\bf Brownian} and {\bf Random}  present 
the expected results for the RDP. The stationary time series {\bf Chaos} presents constant global behavior 
superimposed to a fast oscillation of large amplitude and period of ~14 sampling times. This oscillation 
corresponds to the wandering of the trajectory from one symmetric spiral to the other around the two fixed 
points of the Lorenz attractor. The time series {\bf Noise} and {\bf Random} 
present also a constant RDP but no structure of any 
scale can be identified aside from random  and fast variations of small amplitude. The time series {\bf Brownian} 
presents very large RDP for short times with a very fast decreasing amplitude as time increases, as expected 
from the correlations of such time series. The time series {\bf Pnoise} present a decreasing RDP revealing the 
temporal correlations introduced in this otherwise stationary stochastic time series. The two other real 
world time series {\bf Eeg} and {\bf Ekg} present an overall constant RDP which corresponds to stationarity from the 
point of view of this analysis. In both cases, although more evident in the case of {\bf Eeg}, a smaller 
structure is superimposed without well defined cycles or periods. Some representative RDP of the test time 
series are presented in figure 1.

\begin{figure}[h]

\vspace{-50pt}

\centerline{\hspace{-3.3mm}
\epsfxsize=8cm\epsfbox{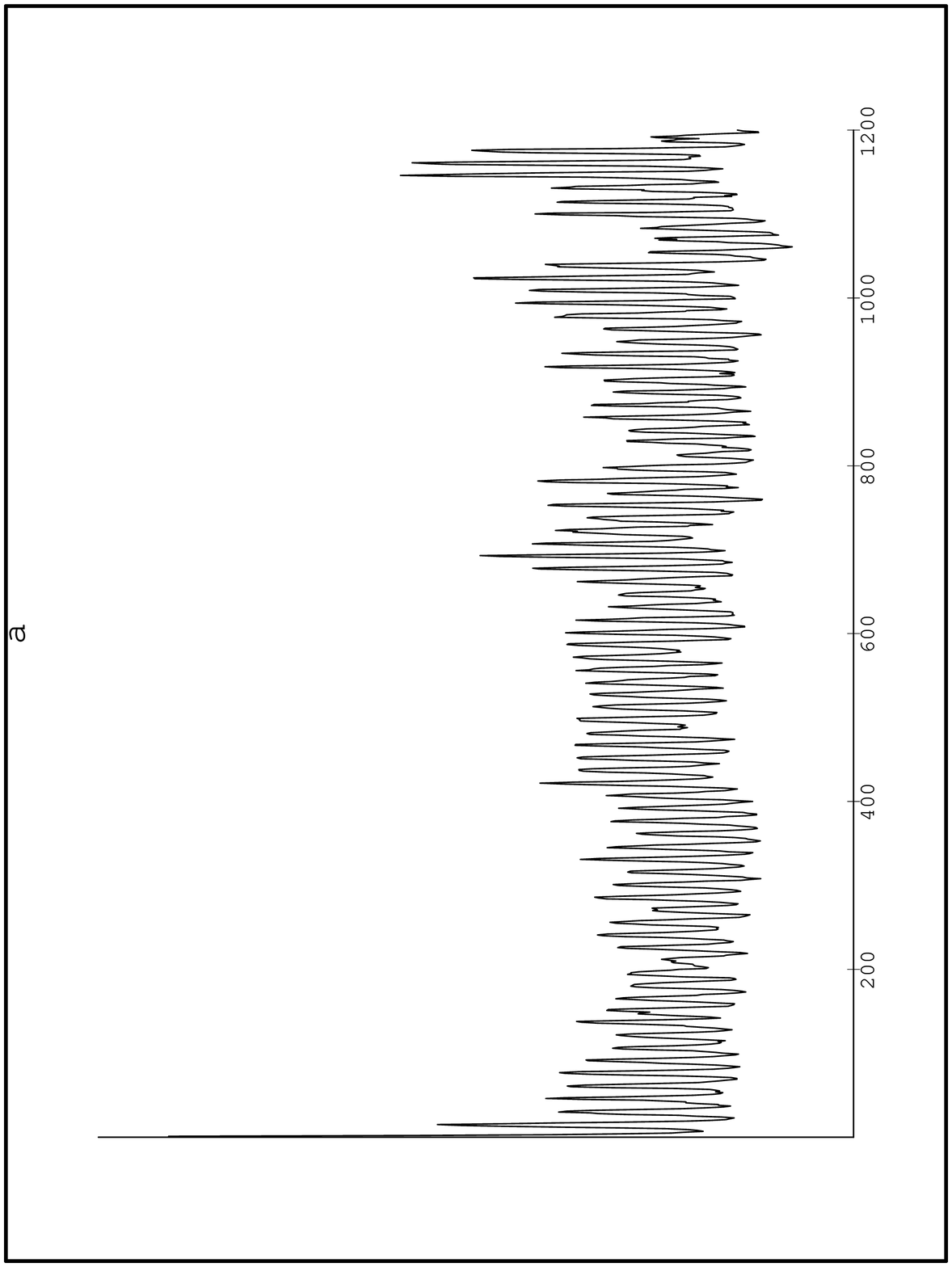}
\hspace{-1cm}
\epsfxsize=8cm\epsfbox{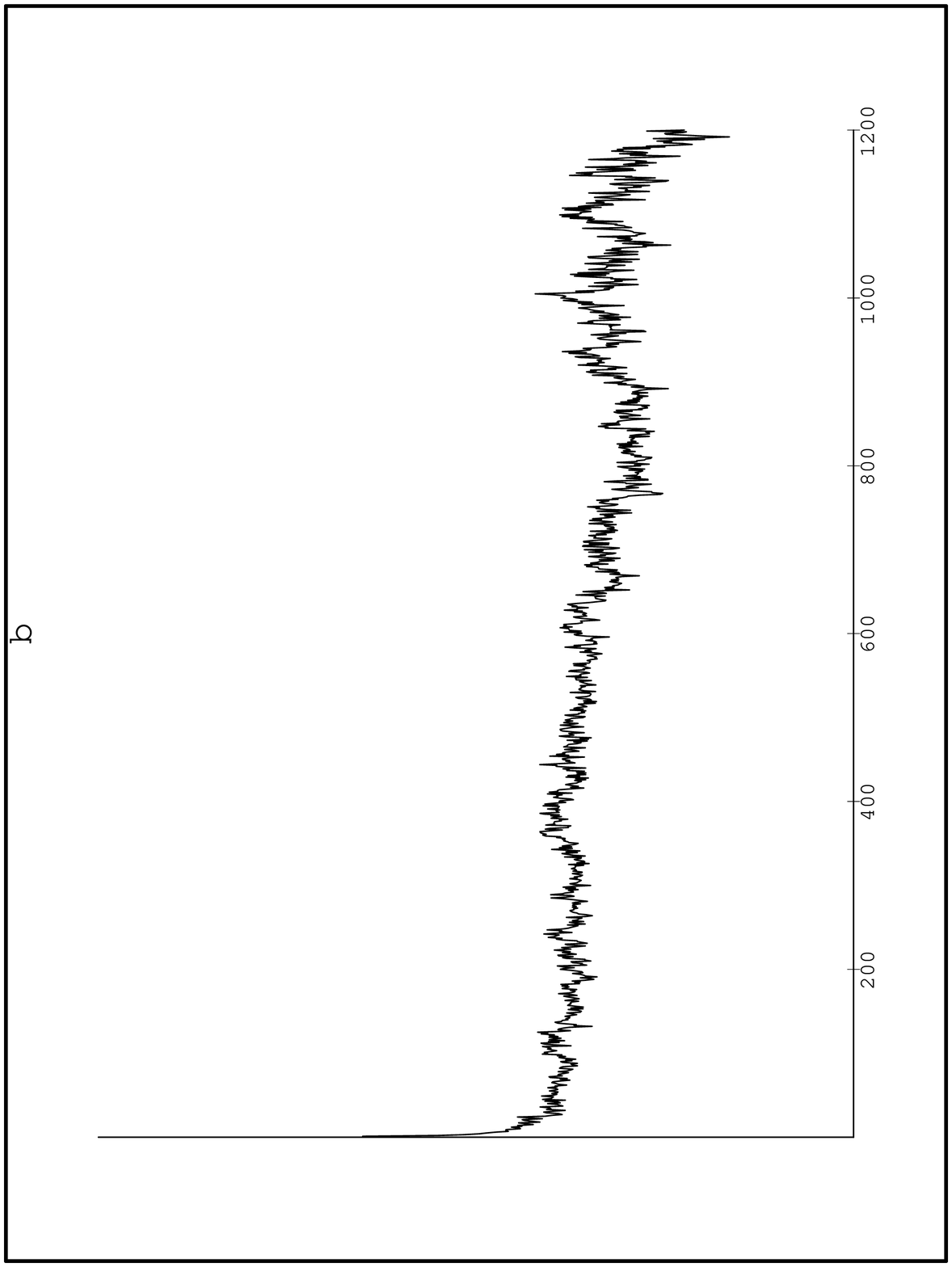}}

\vspace{.0001cm}
\centerline{\hspace{-3.3mm}
\epsfxsize=8cm\epsfbox{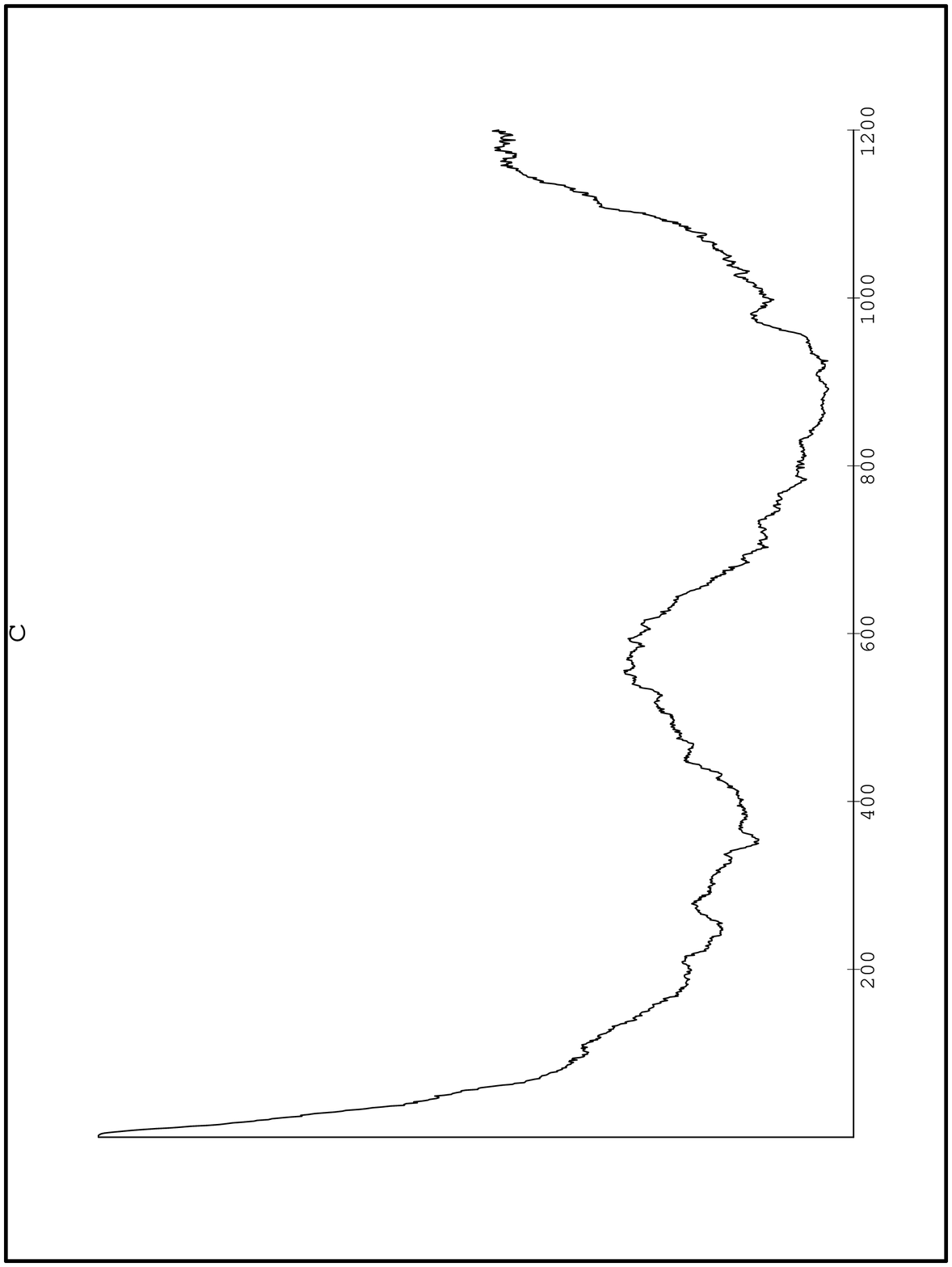}
\hspace{-1cm}
\epsfxsize=8cm\epsfbox{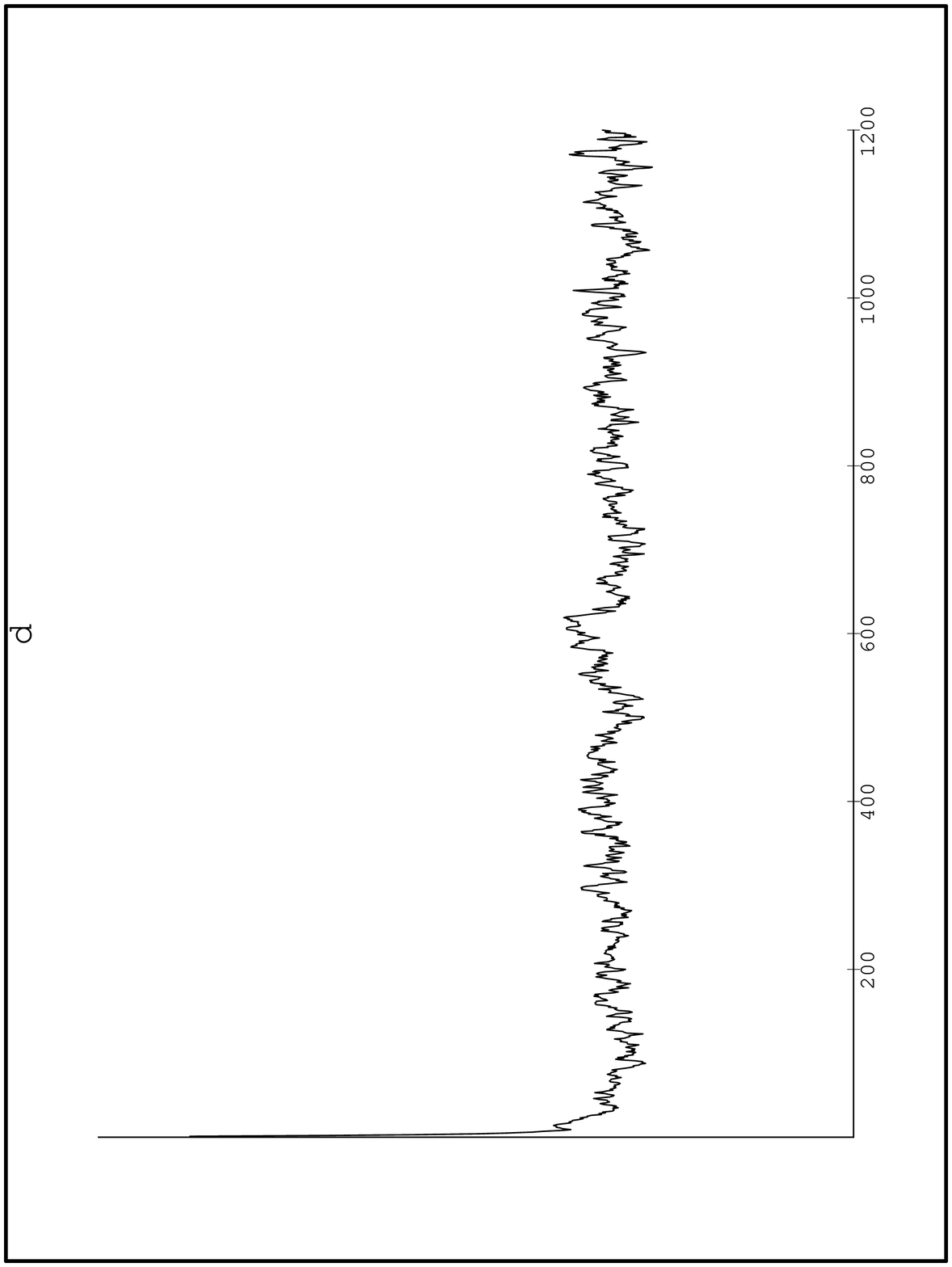}}

\vspace*{-0.3cm}


\caption{ {\footnotesize Recurrence Density Plot (RDP), for some test time series. 
The plot indicates the proportion 
of points encountered at a given spatial distance as a function of the time between such points. 
a) Time series {\bf Chaos}, b) time series {\bf Pnoise}, c) time series 
{\bf Brownian}, and d) time series {\bf Ekg}. 
The time series {\bf Eeg}, {\bf Noise} and {\bf Random}  present similar RDP to 
that of the time series {\bf Ekg} in the sense of an overall constant behavior. However, 
the two stochastic and stationary time series present smaller and more random-like 
amplitude oscillations with a faster initial decreasing behavior.}}

\end{figure}

The RDP for the {\bf Tts} presents very clear nonstationarity with monotonic decreasing behavior superimposed 
to a small structure with no clear periods. The twenty one {\bf TTts} can be divided in three qualitatively similar 
groups. The first group conformed by {\bf TTts} 2,3,5,6,7,8,11,12 and 13 present a stationary-like RDP with some 
superimposed structure more evident in some cases than in others. This small structure is particularly evident 
in {\bf TTts} 6 with a relatively well defined period of $\sim 100$ sampling times or monthly measurements, i.e. between 8 
to 9 years. From this group of  nine apparently stationary time series the short time structure is particularly 
noise-like for the {\bf TTts} 7, 8 and 13. The {\bf TTts} 18 has slow decreasing RDP with a 
strong noise-like small variations. 
The second group is given by {\bf TTts} 4,9,10,19 and 20 which seems clearly nonstationary with minor differences 
compared to the corresponding RDP of the {\bf Tts}. The third and most interesting group is given by 
{\bf TTts} 14,15,16,17,21 and 22. These six time series present good stationarity from the RDP point of view, 
in particular for {\bf TTts} 21. The other five {\bf TTts} present slightly decreasing trends. All six {\bf TTts} of the 
third group present relevant small structure without well defined periods but clearly discriminable from 
noise-like behavior. Representative RDP of the three groups of {\bf TTts} and {\bf Tts} are presented in figure 2.

\begin{figure}[h]

\vspace{-50pt}

\centerline{\hspace{-3.3mm}
\epsfxsize=8cm\epsfbox{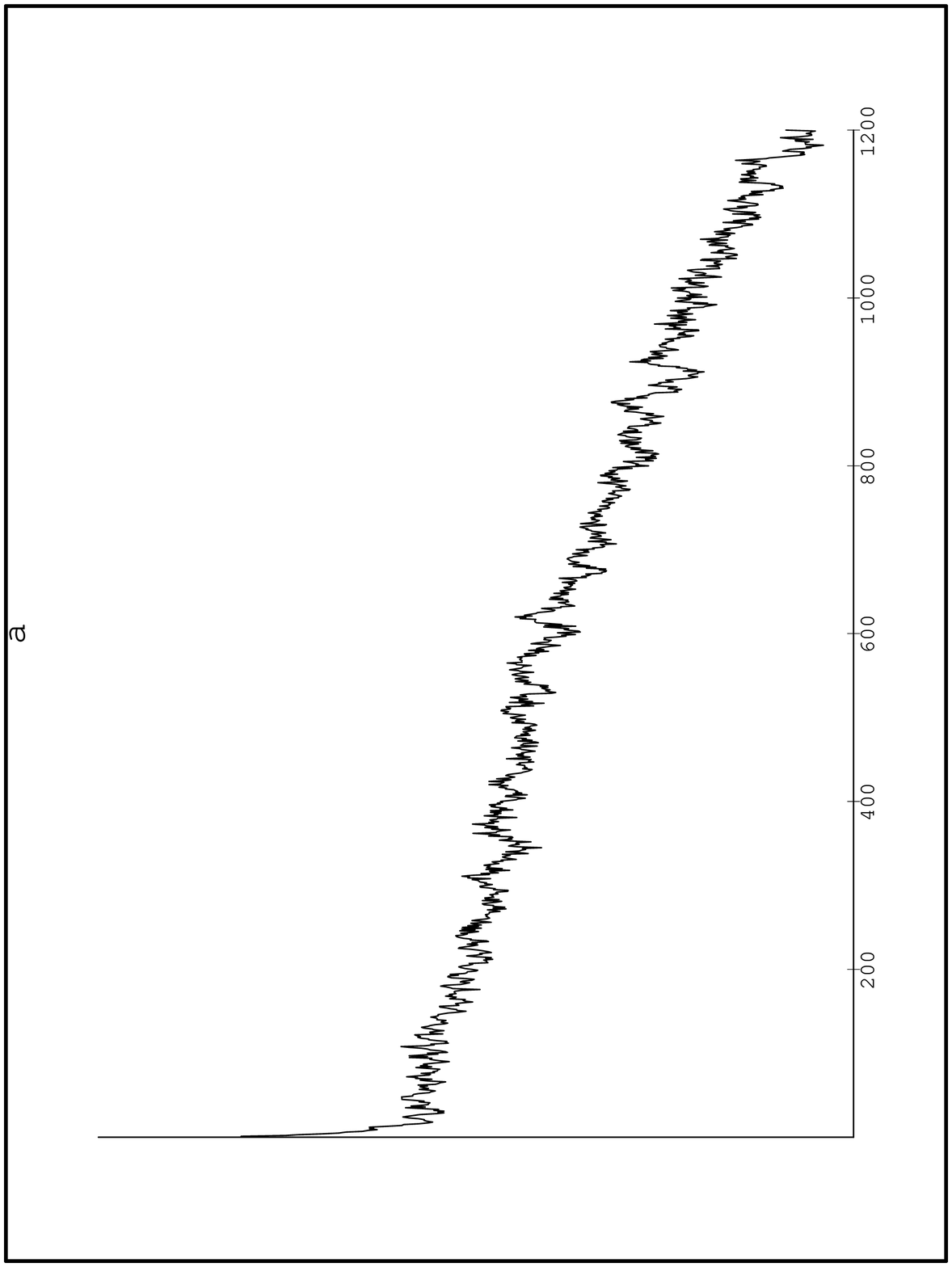}
\hspace{-1cm}
\epsfxsize=8cm\epsfbox{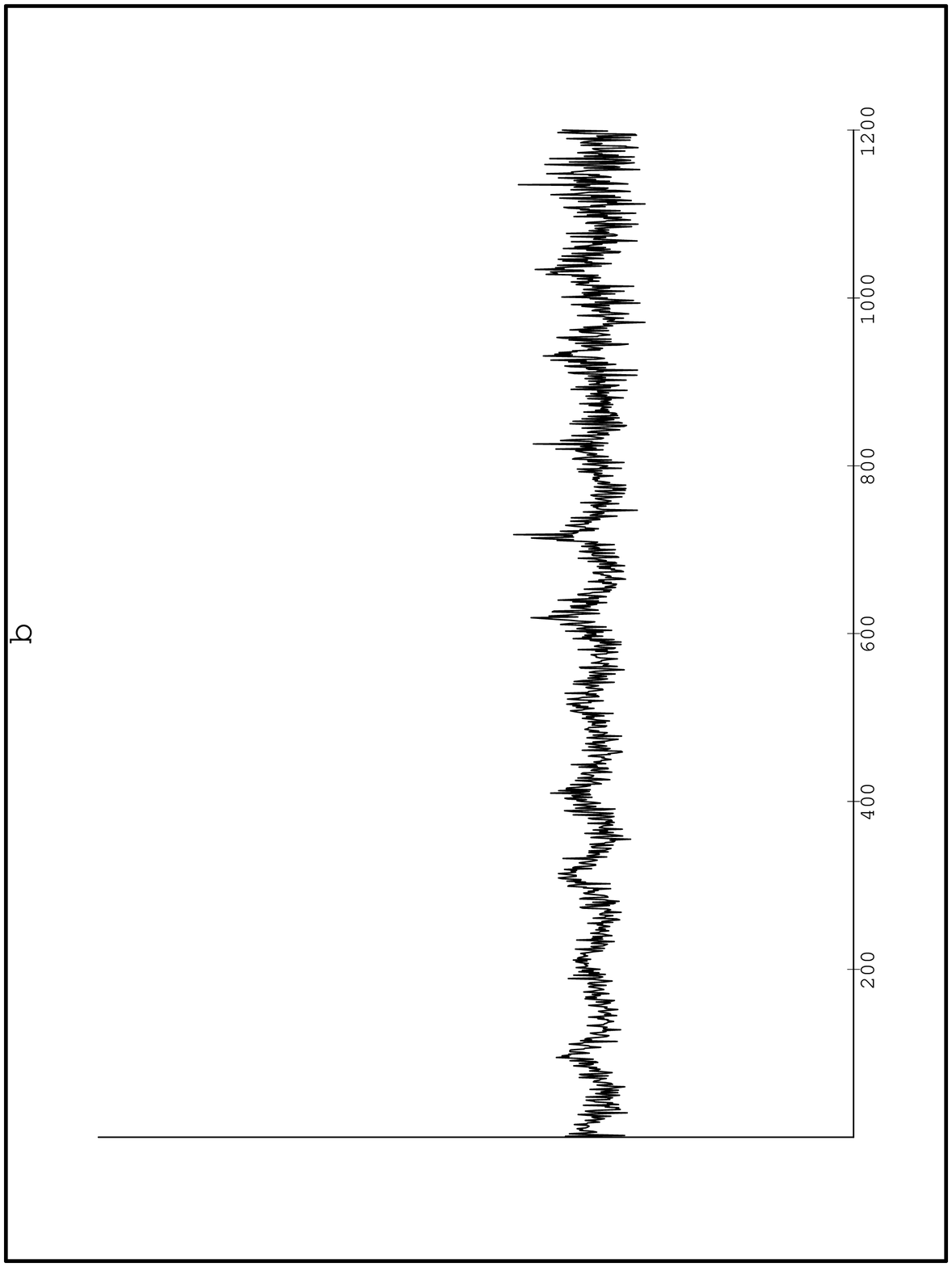}}

\vspace{.0001cm}
\centerline{\hspace{-3.3mm}
\epsfxsize=8cm\epsfbox{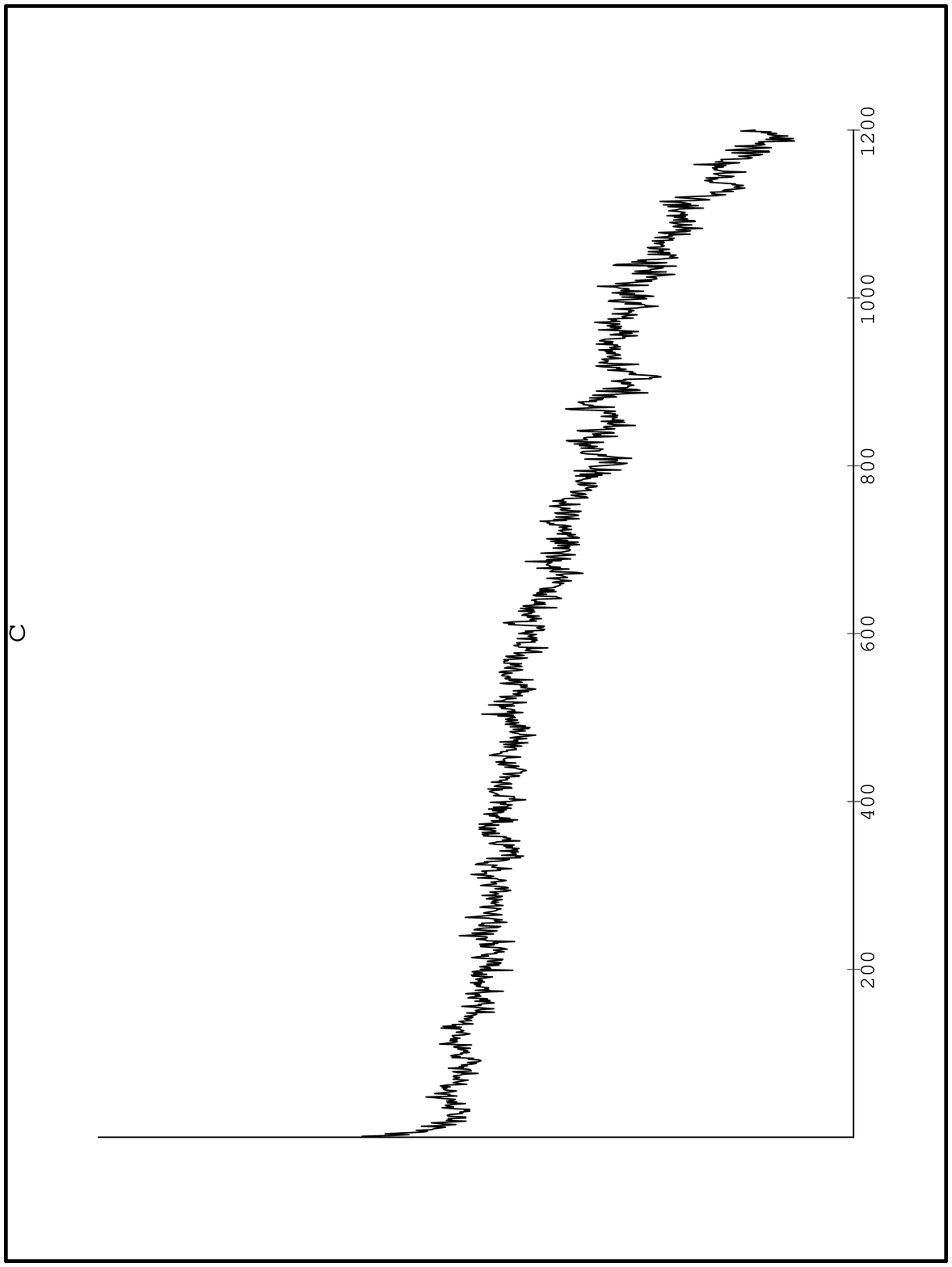}
\hspace{-1cm}
\epsfxsize=8cm\epsfbox{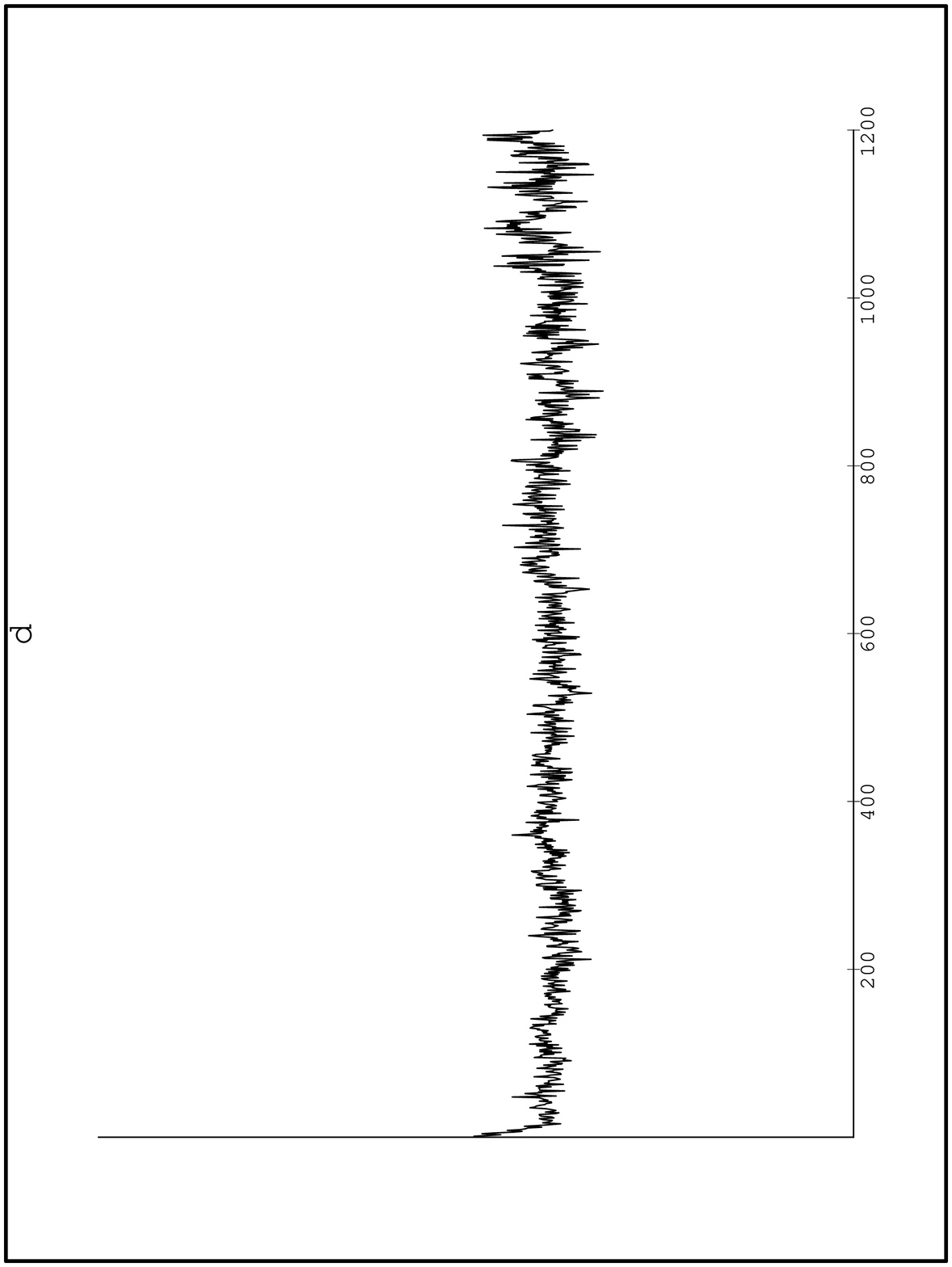}}

\vspace*{-0.3cm}


\caption{ {\footnotesize RDP for {\bf Tts} and some characteristic {\bf TTts}. The plot indicates the proportion 
of points encountered at a given spatial distance as a function of the time between such points. 
a) {\bf Tts}. b) {\bf TTts} 6, 
representative of stochastic-like behavior, i.e. initialy small and  
overall constant density: {\bf TTts} 2, 3, 5, 6, 7, 8, 11, 12 and 13. 
In the particular case of {\bf TTts} 6 certain periodic superimposed structure is 
observed which is not at all the case for the rest of the {\bf TTts} of this group. c) {\bf TTts} 19, 
representative of clearly nonstationary RDP: {\bf TTts} 4, 9, 10, 19 and 20. d) 
{\bf TTts} 21, representative of nonstochastic stationary RDP, e.i. initially large and decreasing 
density followed by an overall constant density:
{\bf TTts} 14, 15, 16, 17, 20, 21 and 22. The short but not 
negligeable initial decreasing behavior of RDP, characteristic of nonstochastic time series, 
is not clearly observed on the RDP since we are mainly interested on  
the long term overall behavior of RDP needed to assess 
the stationarity of the time series.}}

\end{figure}

\underline{Nonlinear Cross Prediction Error} (CPE): Nonstationarity in a deterministic time series may 
be identified by  overall trends in the  CPE, as the predicted and predictee segments correspond to more and more distant parts of 
the time series. The time series {\bf Brownian} shows the largest range of values and variability of CPE. For the 
time series {\bf Noise} and {\bf Random}, we get large CPE with a homogeneous random  appearance.  These two 
characteristics of CPE are representative of uncorrelated stochastic time series because it is 
irrelevant which segments are used as predictee and predicted. In the case of the time series 
{\bf Pnoise} and {\bf Brownian} CPE is in general large with clear patterns of valleys and crests indicating 
the corresponding correlations, which are more remarkable in the second case. For the time series 
{\bf Chaos} CPE is in general small with regular patterns indicating permanent correlations due to 
stationary determinism. Increasing trends indicating parameter drifts or changes in the form of 
the attractor are not observed in the time series {\bf Chaos}, as expected. In all cases the results are 
not qualitatively affected by the values of the embedding dimension from 1 to 6. The real world 
time series {\bf Eeg} and {\bf Ekg} present the deterministic characteristic of patterns with general small 
CPE and do not present clear evidence of nonstationarity like increasing trends as predicted 
and predictee segments get further apart. However, for these two time series the results 
are much more sensitive (than the other test 
time series), to the parameter $\epsilon$ and the uncorrelated stochastic appearance is easily 
obtained. For {\bf Tts} the results are also very sensitive on the parameter $\epsilon$ although, strong 
indications of the mentioned deterministic characteristics persist. The CPE plot for {\bf Tts} presents 
four characteristic regions determined by the two halves of the {\bf Tts} when we use them as a prediction 
or to be predicted alternatively. The first half is not good for prediction and similarly bad 
predictions of it are obtained. The second half is good to make predictions and good predictions (small CPE) 
of it are also obtained apart from the last two or three segments which present larger CPE. 
These are consistent results with the fact that the first half of {\bf Tts} is more 
contaminated than the second, and the last part of the second half may present strong trends. These results are 
also consistent with those of nonlinear statistics which give indications of less determinism for the 
first half of {\bf Tts}.

For all time series studied with 1356 data points, we have used 20 nonoverlapping sections which 
corresponds to $l\sim 67$ data points. For the time series {\bf Chaos} $l\sim 67$ 
data points is a time length larger than the ``coherence time" of the Lorenz attractor, $\sim 14$ sampling times 
identified by the RDP. For the rest of the studied time series  $l\sim 67$ data points  
seems to be a good compromise between the sampling time and the length of the time series.
Similarly to {\bf Tts}, the two groups {\bf TTts} 4, 9, 10,16, 17 and 19, and {\bf TTts} 14, 15, 18, 20, 21 and 22 present 
relatively large ranges of CPE with clearly small regions. The second group looks slightly more 
stationary with less trends and is less sensitive to $\epsilon$. The remaining {\bf TTts} 2, 3, 5, 6, 7, 8, 11, 12 
and 13 give CPE with homogeneous random  variability with no patterns and no predominantly small regions, 
i.e. uncorrelated stochastic-like behavior.
From the point of view of the CPE, several of the treatments applied to {\bf Tts} to isolate the deterministic 
information produces the contrary effect and noise-like time series are obtained as mentioned above. When 
this was not the case, the apparently nonstationarity of {\bf Tts} was in some cases reduced, in particular for 
{\bf TTts} 9, 10, 14 to 22. In  figures 3 and 4 we present representative CPE 
plots for {\bf Tts}, some test time series and some {\bf TTts}.

\begin{figure}[h]

\vspace{-50pt}

\centerline{\hspace{-3.3mm}
\epsfxsize=8cm\epsfbox{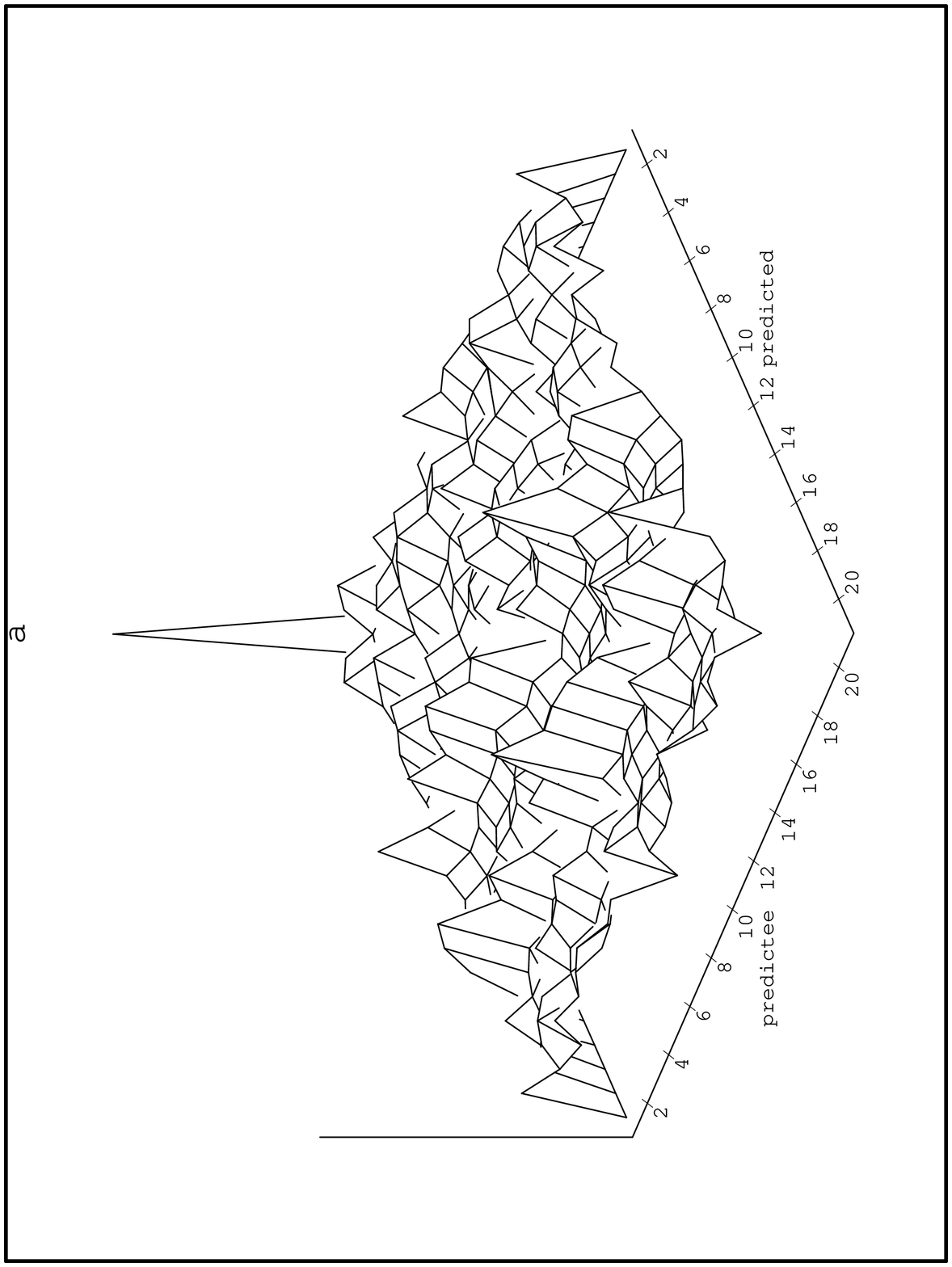}
\hspace{-1cm}
\epsfxsize=8cm\epsfbox{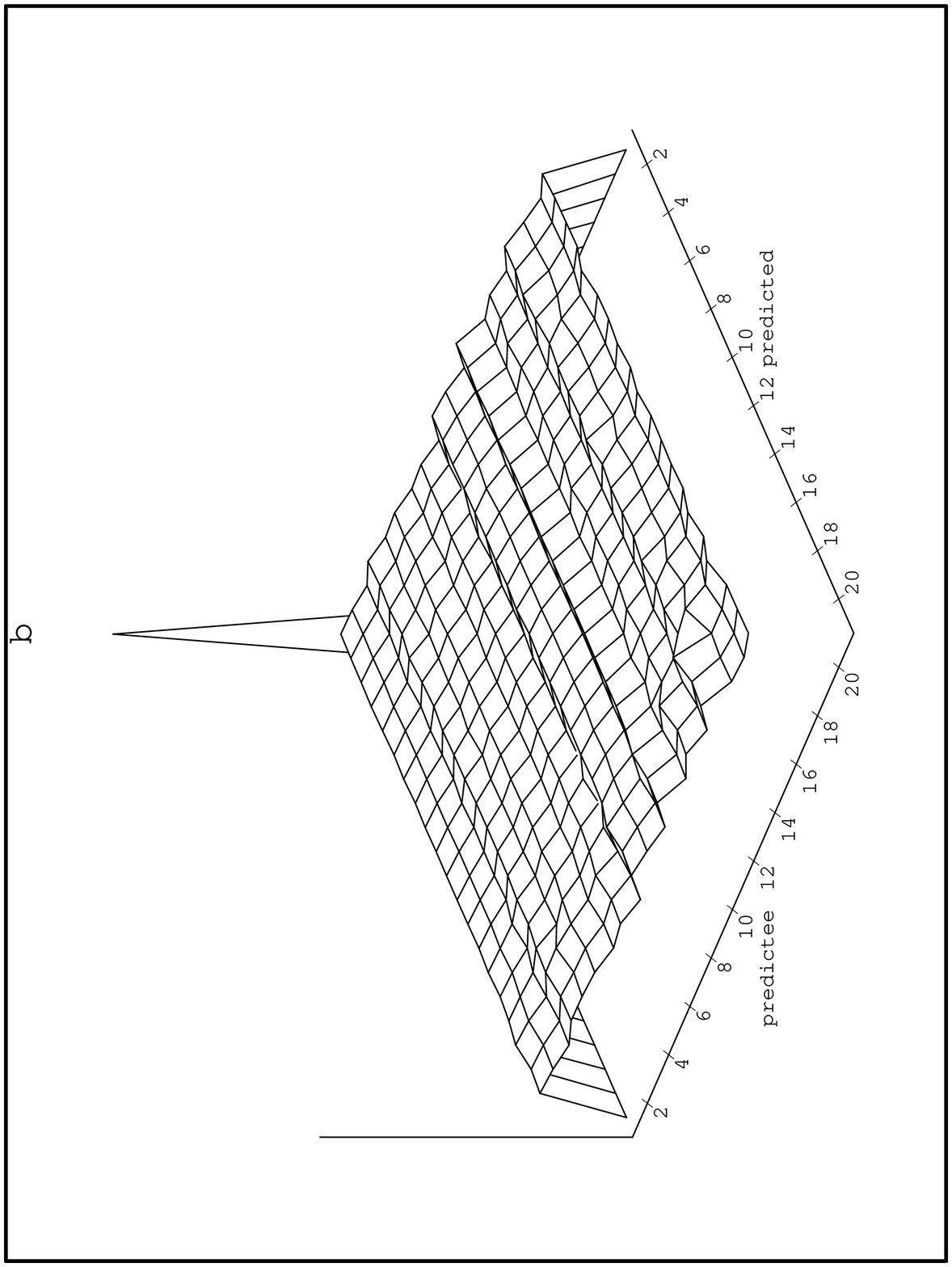}}

\vspace{.0001cm}
\centerline{\hspace{-3.3mm}
\epsfxsize=8cm\epsfbox{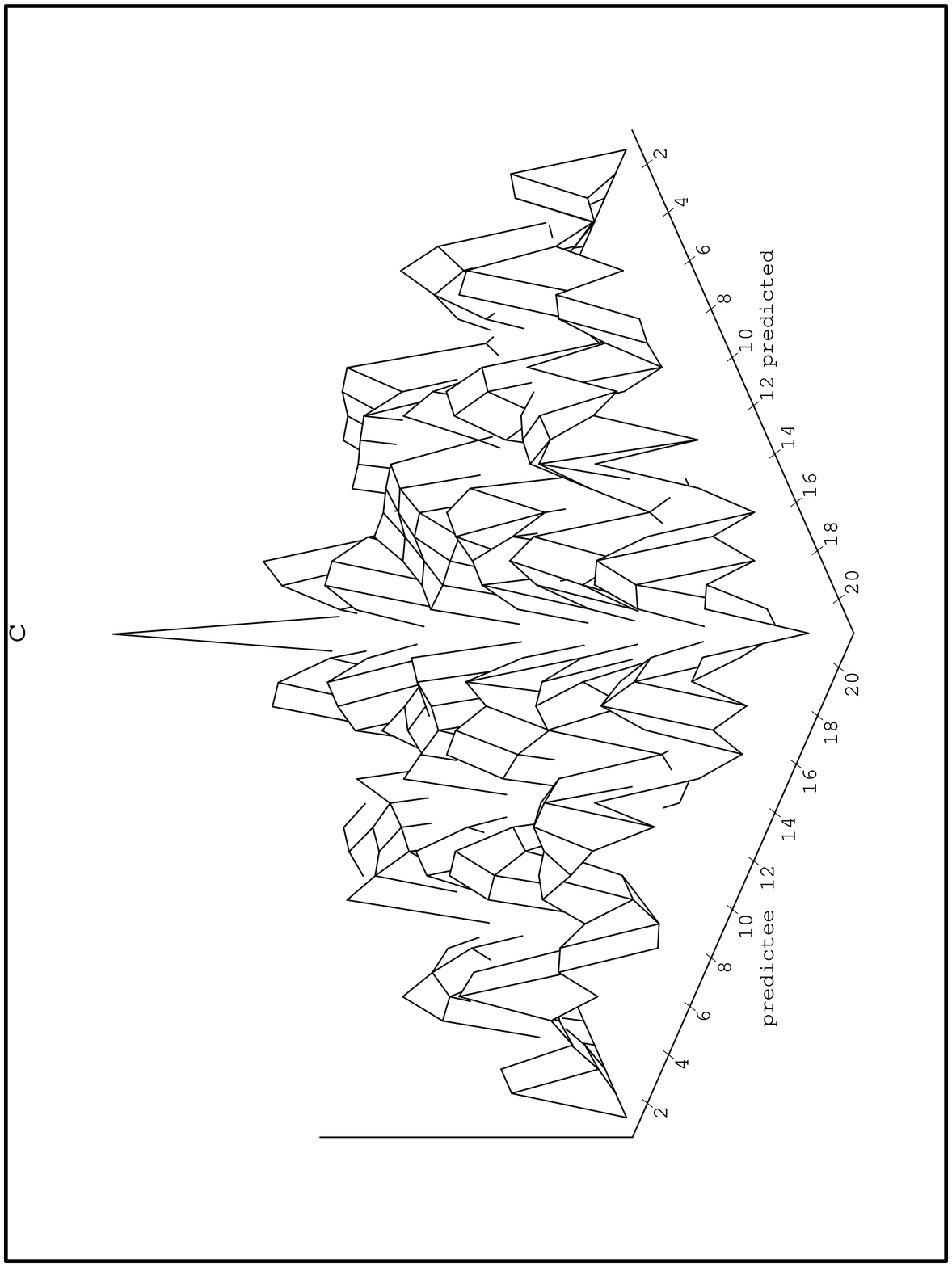}
\hspace{-1cm}
\epsfxsize=8cm\epsfbox{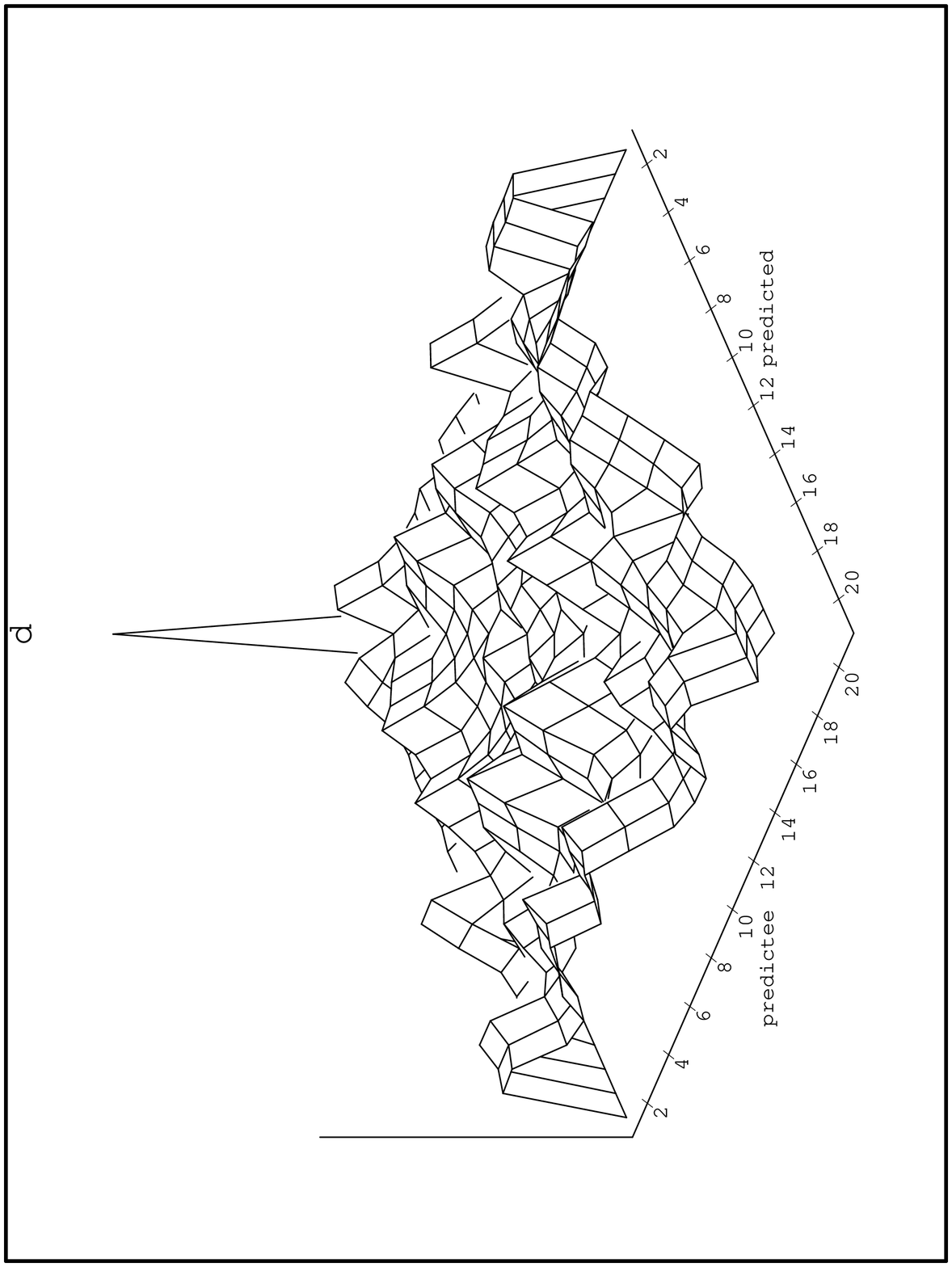}}

\vspace*{-0.3cm}


\caption{ {\footnotesize Cross-Prediction Errors, (CPE). The two horizontal axes indicate the 
index of the segments used to predict and to be predicted. The CPE values are given 
in arbitrary units proportional to one half of the range of values of the 
corresponding time series. For clarity, we have fixed CPE to the maximum possible value (one half of the 
range of the time series)  for the point $(0,0)$ and to 
zero for the rest of the pairs $(predicted,0)$ and $(0,predictee)$. 
a) CPE plot for time series {\bf Chaos}, b) CPE plot 
for time series {\bf Noise}, c) CPE plot for time series {\bf Brownian}, and 
d) CPE plot for time series {\bf Pnoise}.}}

\end{figure}

\begin{figure}[h]

\vspace{-50pt}

\centerline{\hspace{-3.3mm}
\epsfxsize=8cm\epsfbox{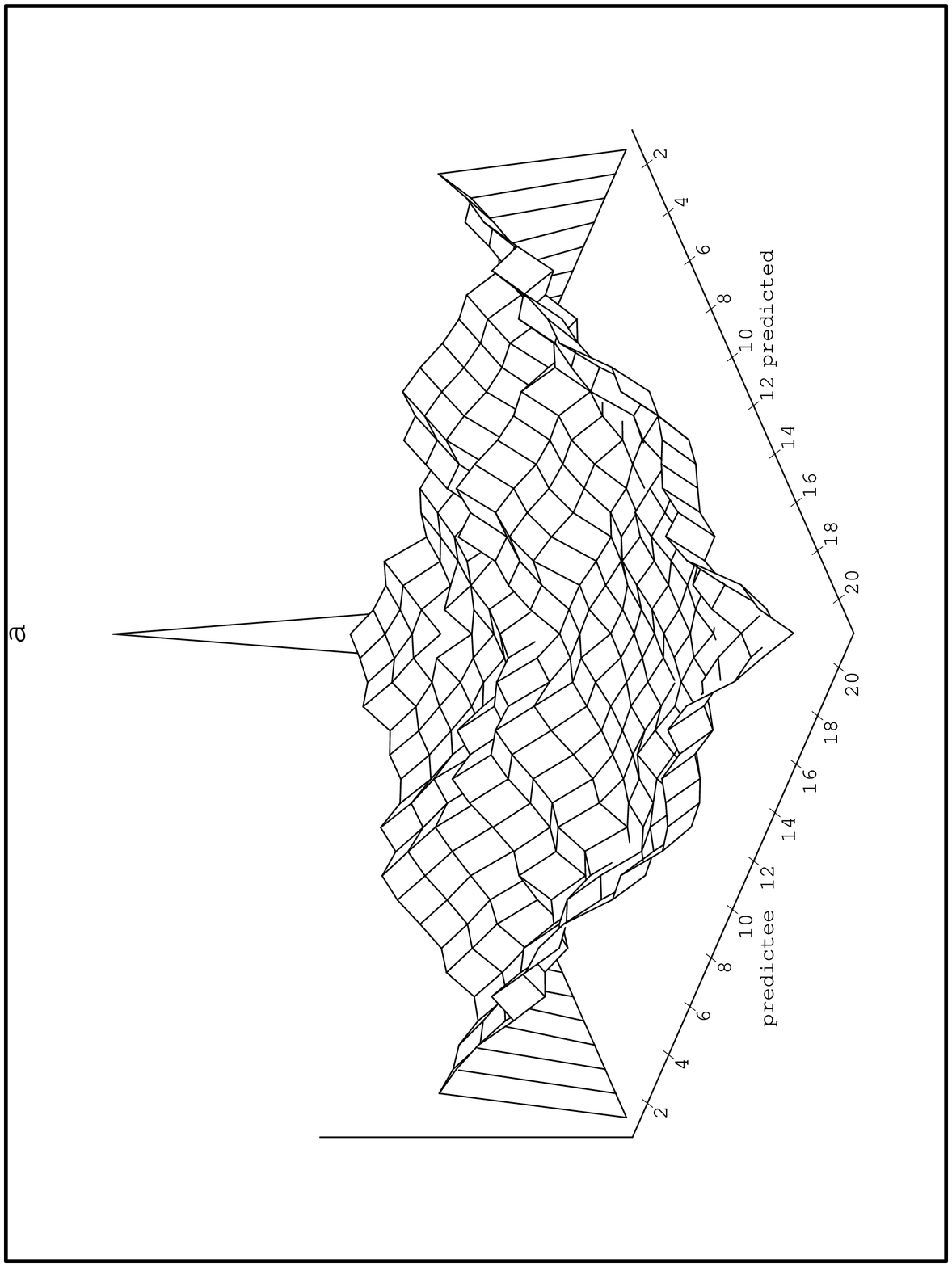}
\hspace{-1cm}
\epsfxsize=8cm\epsfbox{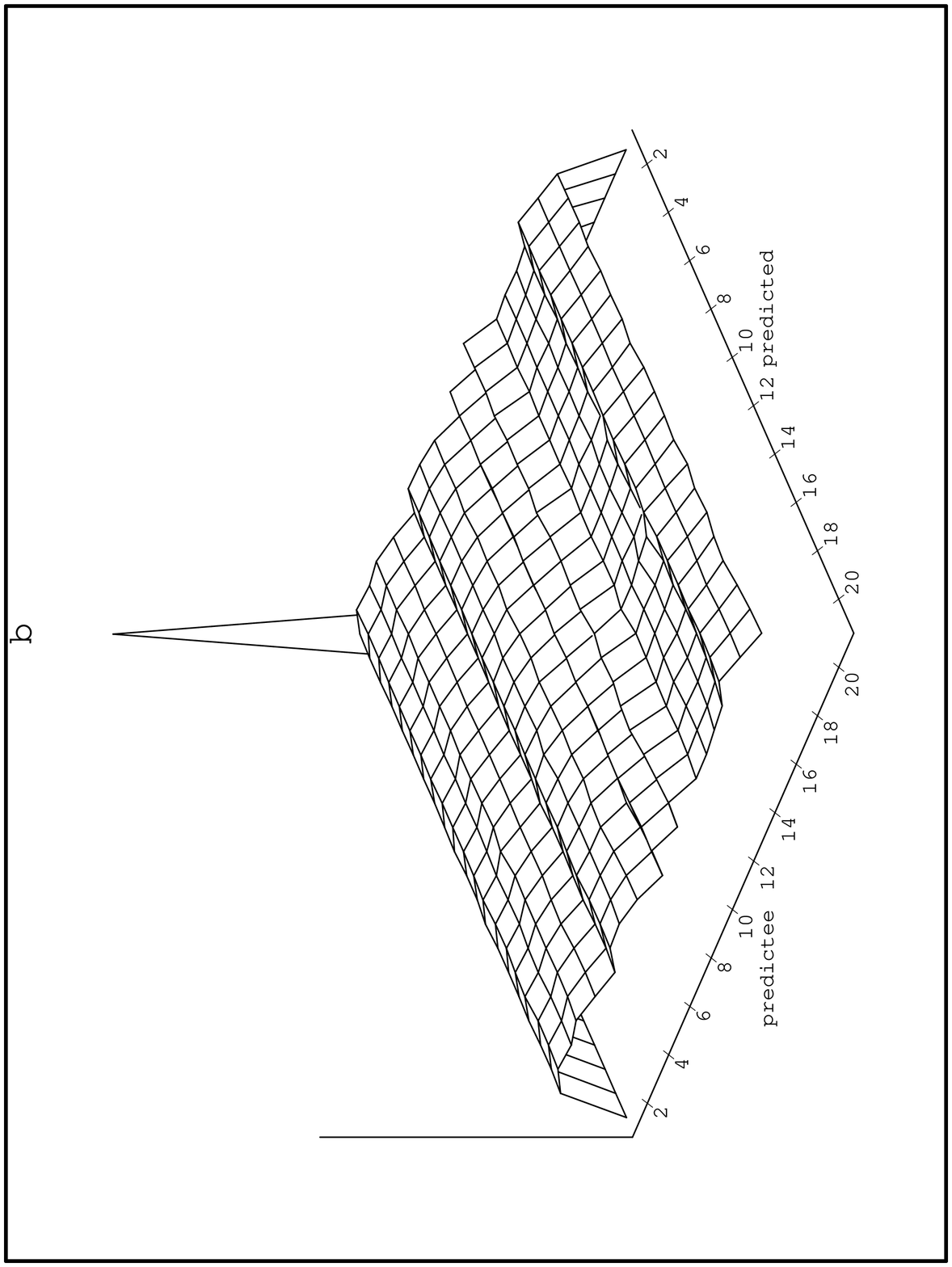}}

\vspace{.0001cm}
\centerline{\hspace{-3.3mm}
\epsfxsize=8cm\epsfbox{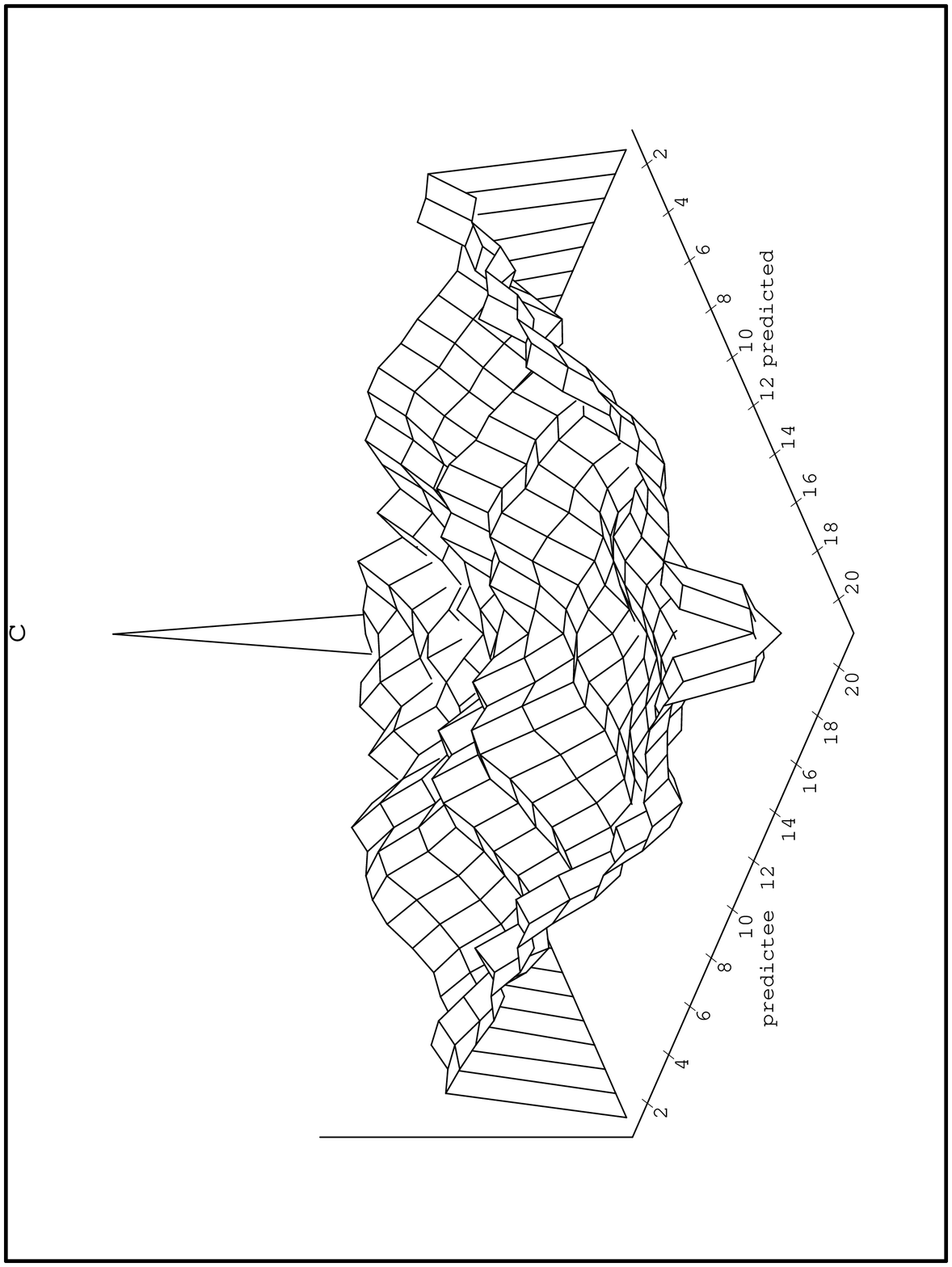}
\hspace{-1cm}
\epsfxsize=8cm\epsfbox{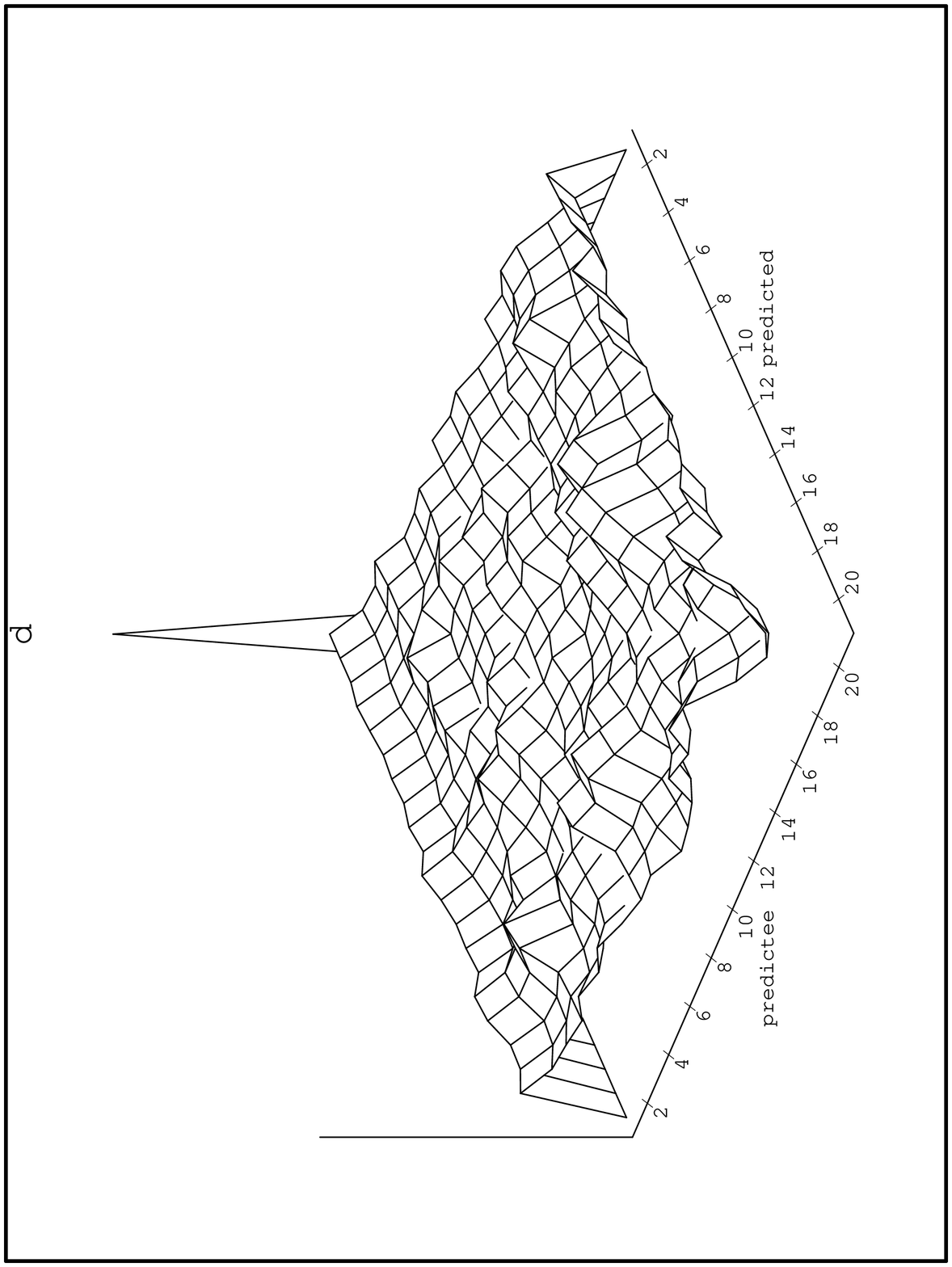}}

\vspace*{-0.3cm}


\caption{{\footnotesize CPE plots for {\bf Tts} and some representative {\bf TTts}. 
For interpretation of the plots see figure 3. a) 
CPE plot for {\bf Tts}. b) CPE plot for {\bf TTts} 11 as representative of the uncorrelated and stochastic-like 
{\bf TTts} 2, 3, 5, 6, 7, 8, 11, 12 and 13. c) CPE plot for {\bf TTts} 4 as a representative of  {\bf TTts} 4, 9, 10, 16, 17 
and 19 which remain similar to {\bf Tts} from the point of view of CPE. d) CPE plot for {\bf TTts} 14, 15, 18, 20, 21 
and 22 which become more stationary from the point of view of this method.  }}

\end{figure}

For all time series studied in this work the sensitivity on $\epsilon$  seems to depend  strongly 
on the complexity of the system as well as on the characteristics of the time series. As more deterministic 
or correlated  and contaminated is the time series more sensitive to $\epsilon$ it is and more easily 
it gives uncorrelated stochastic-like results. However, $\epsilon =40 \% $ of the range of the time 
series, works in general well and allows as to compare all of the time series on equal footing. The 
results of CPE were in all cases weakly dependent on small values of $m$, between 1 and 3, when  
$\tau =1$  \cite{Nota1}.

\underline{Space Time Separation Plot} (STSP): The stationary and deterministic 
time series {\bf Chaos} numerically 
generated with well known geometrical 
and dynamical characteristics, using an appropriated sampling time and good resolution of eight digits 
data points permits to sufficiently sample the attractor avoiding large temporal correlations. For the time 
series {\bf Chaos} all the curves of the STPS corresponding to different percentages, start at zero and then a 
regular oscillation, without any trend, is observed for increasing time. Since the cycle length of $\sim 14$ 
sampling times, is much smaller than the observation period, 1356 sampling times, the oscillation is 
harmless. We are interested in the first time steps where the plots increase consistently revealing 
the length of the temporal correlations \cite{PSVM92, KS97b}. 
The time series {\bf Noise}, {\bf Random}  and {\bf Pnoise} show the expected 
saturated behavior corresponding to stochastic 
uncorrelated time series. In these last three cases the STSP saturates very fast and no other relevant 
behavior apart from numerical oscillations is observed. The only observable difference is for the 
STSP of the time series {\bf Pnoise} with a slightly slower saturation behavior which corresponds to the 
short time correlations involved in such time series. The STSP for the time series {\bf Brownian} present 
small numerical variations and no saturation at any time is obtained representing the 
long term correlations of this time series. The results for the time series {\bf Chaos} are qualitative 
invariant under changes of the embedding parameters for $m=1$ to 7 with $\tau =1$, for $m=1$ to $4$, 
for $\tau =2$ to $4$, 
and a wide range of distances used to define neighbor points.

The STSP analysis of the time series {\bf Eeg} and {\bf Ekg}  does not show clearly the otherwise expected 
nonstationarity because the sampling time and time span of these time series have been controlled 
for this purpose. The corresponding plots saturate relatively fast but slower than the time series 
{\bf Noise} and {\bf Pnoise} representing certain correlation present in these real world time series. 
The STSP for the {\bf Tts} shows good stationarity after a correlation on the order of 
$\sim 10$ to $15$ sampling times, 
corresponding approximately to one year. The relatively long time between measurements and low 
resolution of the {\bf Tts} may be important reasons for such short time correlations. 
The results of the STSP for the {\bf Tts} are also robust with respect to different embedding dimensions and 
small values of the time delay $\tau$. For $ \tau \geq 3$ the results are strongly altered which is expected 
for these sparse time series because  the time scales are importantly changed  
when $\tau$ becomes large. For all {\bf TTts} the results of 
STSP are similar to those of {\bf Tts}. In the cases where {\bf TTts} were obtained by treatments which explicitly damp 
high frequencies or short term correlations as {\bf TTts} 3,7,8,11,12 and 13, STSP presents faster saturation 
although, those STSP are difficult to differentiate from those corresponding to uncorrelated stochastic 
time series.  In figures 5 and 6 we present some representative STSP and 
the approximately saturation times of all 
the studied time series are presented in table 1 column 9. Since we are mostly interested 
in the qualitative behavior, we chose a set of parameter values around which 
the results of STSP are robust to parameter variations and represent well the 
characteristics of the corresponding time series. We can then compare the results 
of the different time series on similar footing. For figures 5 and 6 we have fixed 
$\tau= 1$, $m=1$ and $1/10$ of the maximum number of spatial partitions that each time 
series supports (given by the range of the data points divided by their resolution), 
as the spatial partition used to select near neighbors.

\begin{figure}[h]

\vspace{-50pt}

\centerline{\hspace{-3.3mm}
\epsfxsize=8cm\epsfbox{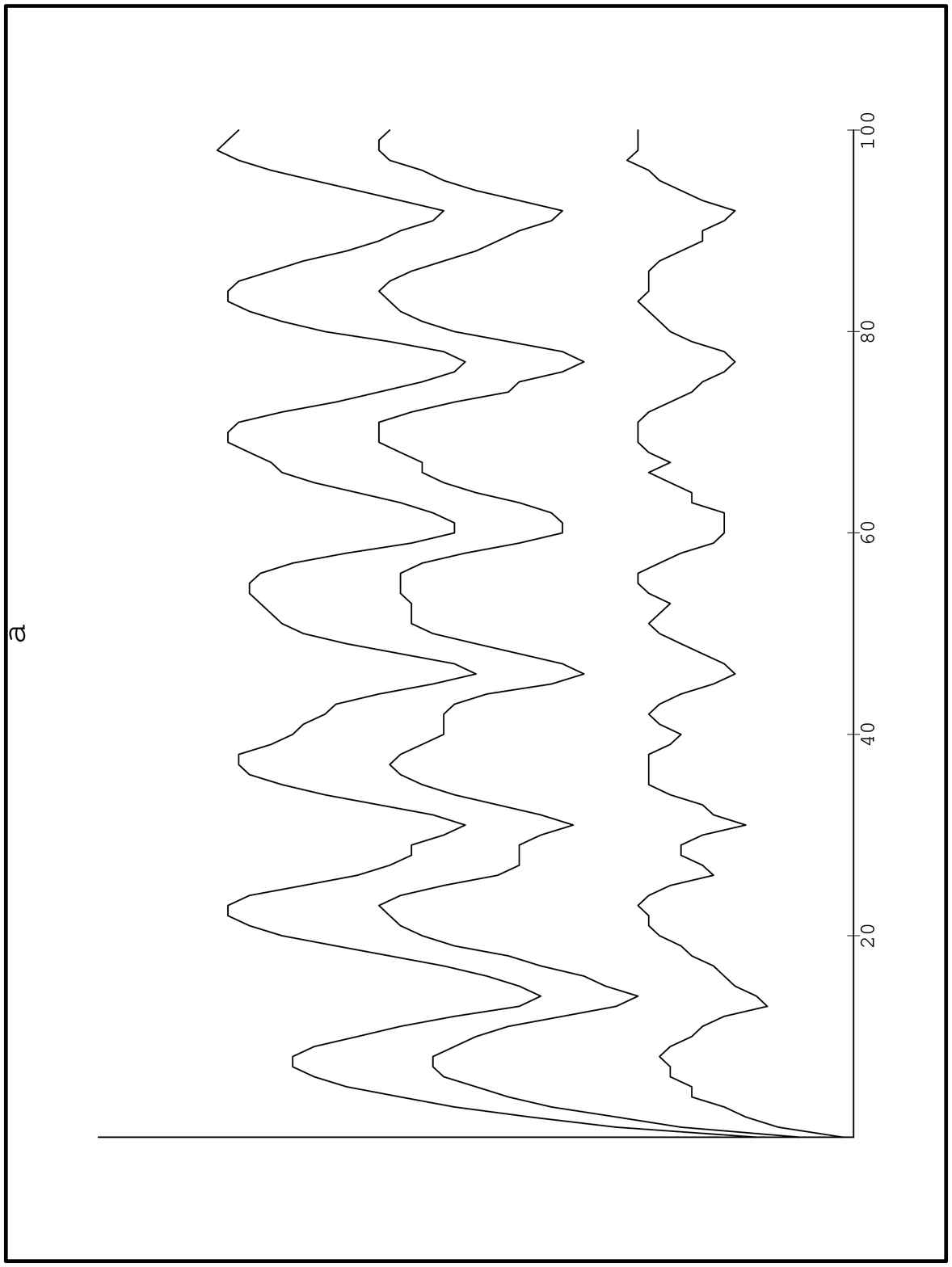}
\hspace{-1cm}
\epsfxsize=8cm\epsfbox{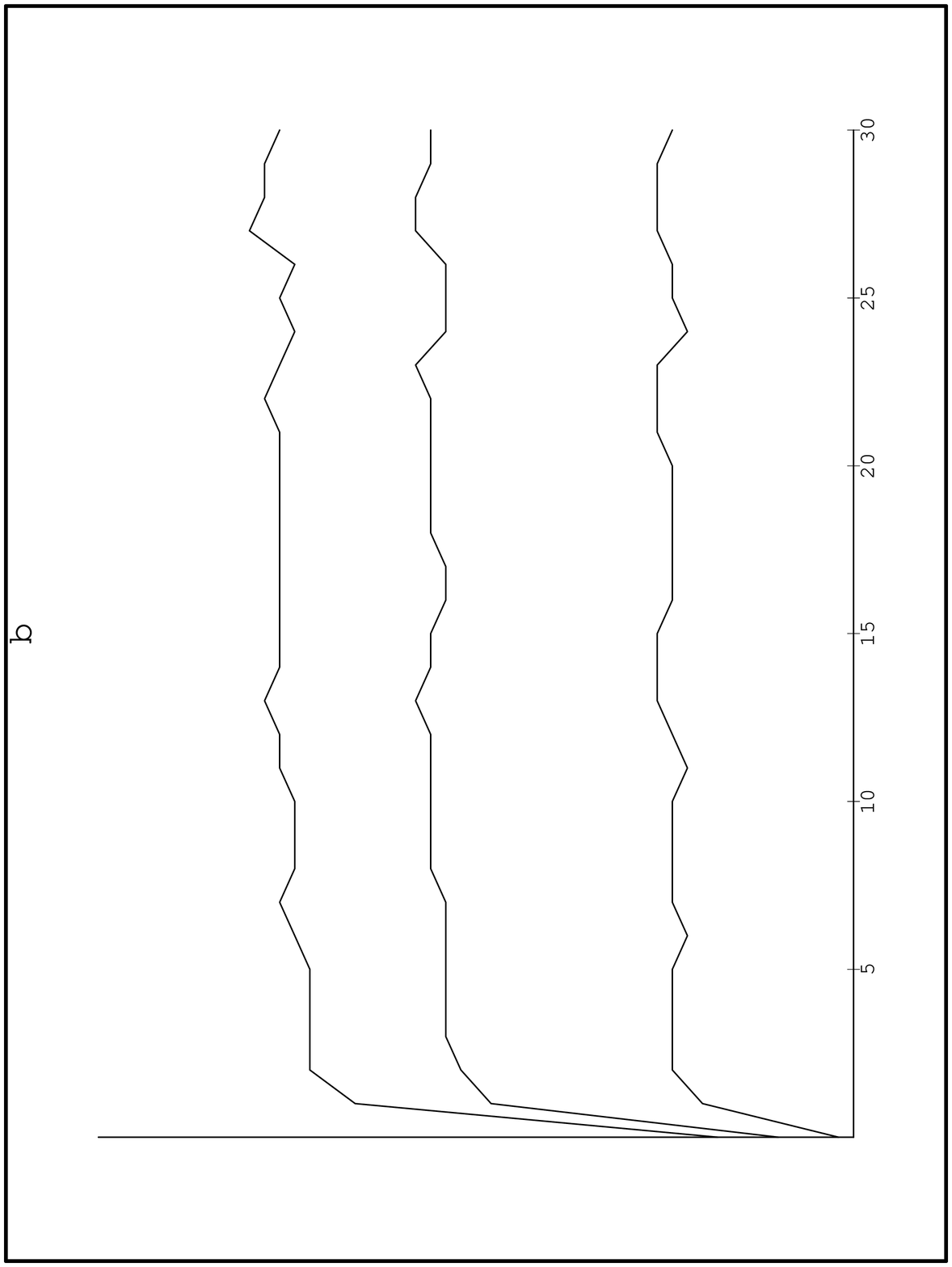}}

\vspace{.0001cm}
\centerline{\hspace{-3.3mm}
\epsfxsize=8cm\epsfbox{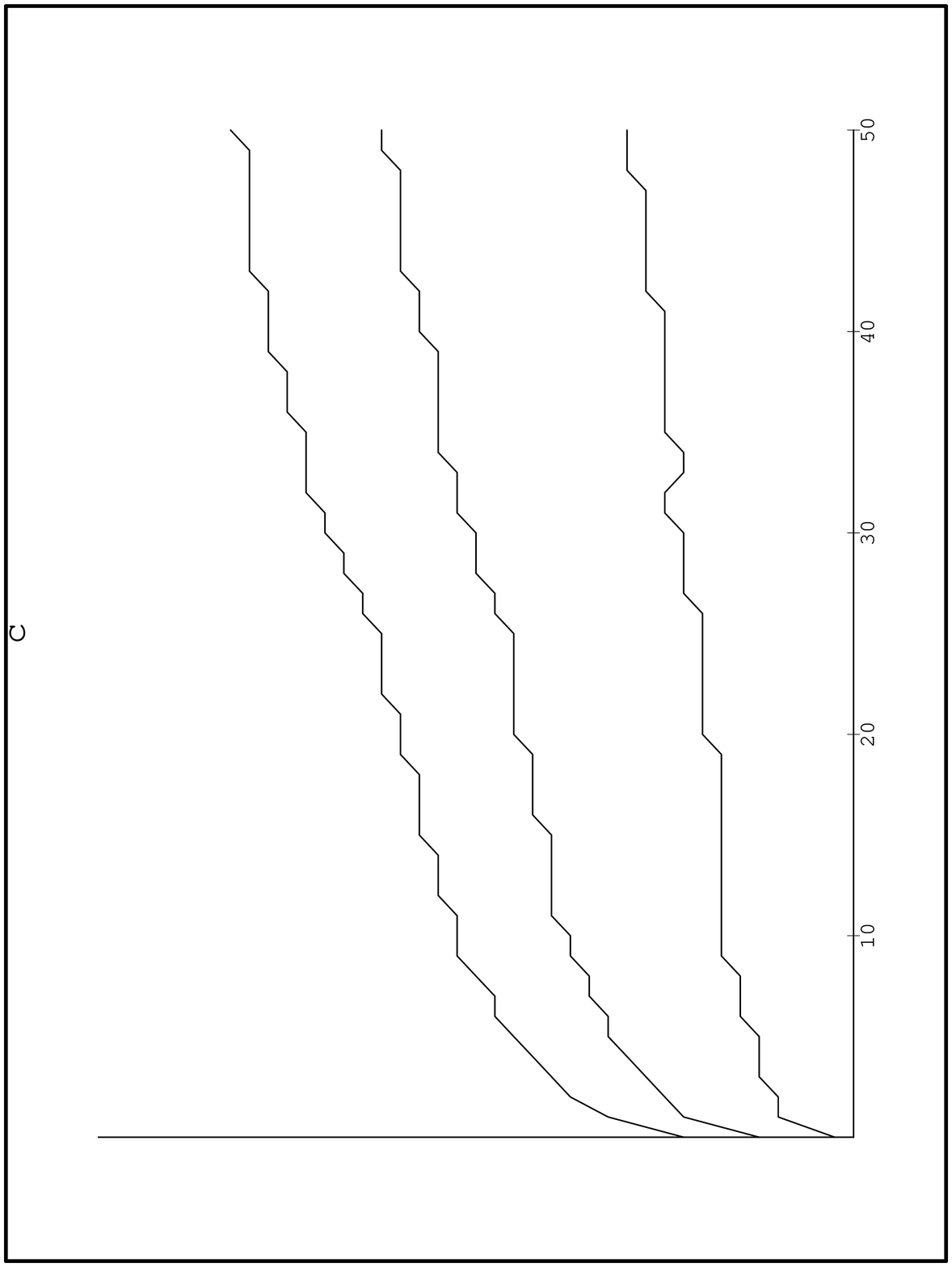}
\hspace{-1cm}
\epsfxsize=8cm\epsfbox{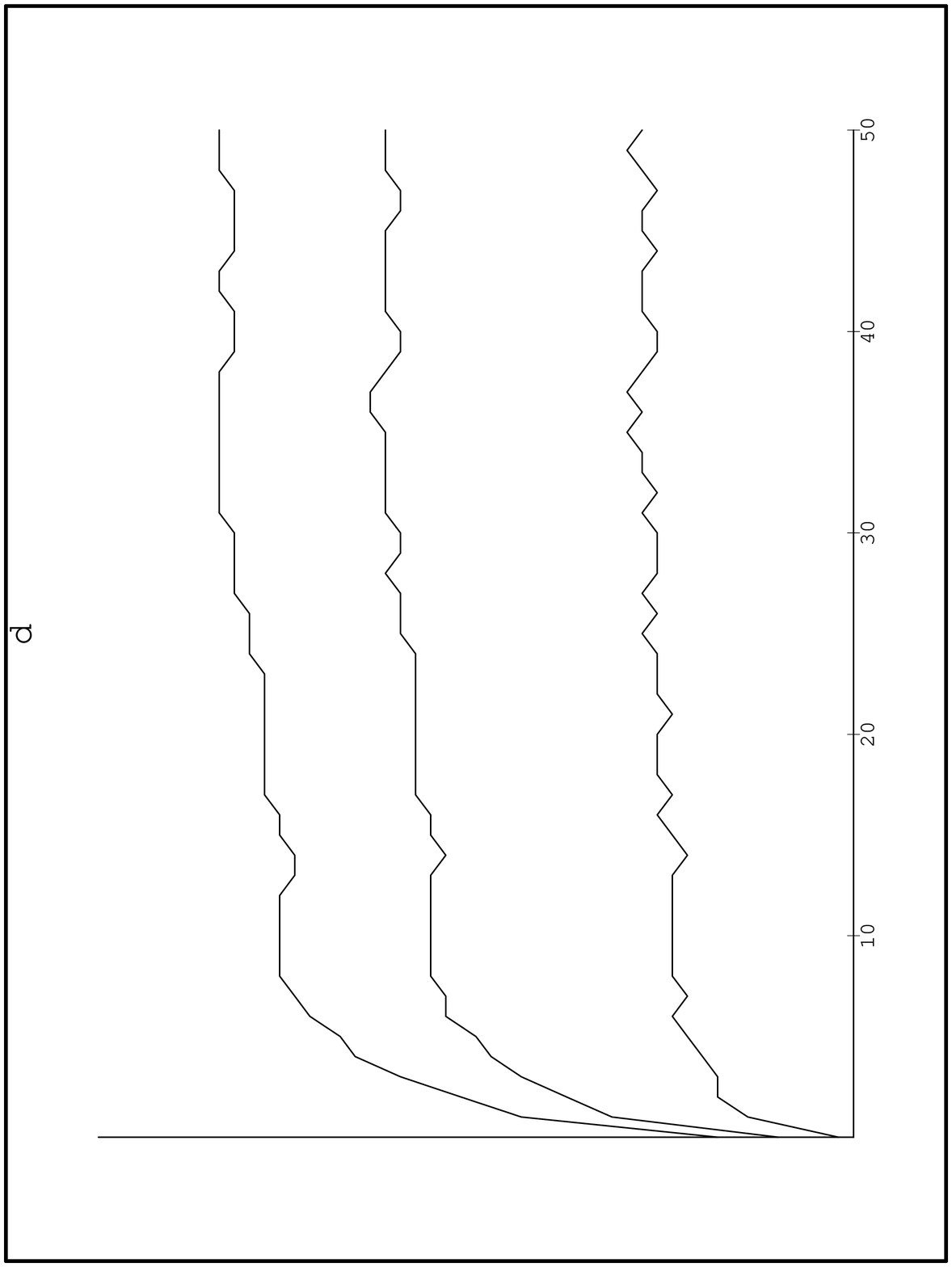}}

\vspace*{-0.3cm}


\caption{ {\footnotesize STSP for some test time series. 
The curves indicate the spatial distance 
obtained as a function of the temporal separation in order to obtain a fixed fraction 
of pairs of all the possible pairs. The three curves 
in each plot represent respectively, 
from bottom to top, 10, 50 and 90\% of all possible pairs of points. 
Each curve was calculated up to 300 or more 
sampling times to make sure of the overall saturation when present. We show 
only the first 100 or less sampling times, as indicated by the horizontal axis, in 
order to clarify the behavior before saturation. a) Time series {\bf Chaos}. 
b) Time series {\bf Pnoise} which present qualitative equal results to 
the time series {\bf Random}  and {\bf Noise} but with a slightly 
slower saturation. c) Time series {\bf Brownian}, the only time series 
which in practice does not present saturation. In practice the 
saturation time appears before the end of the time series because of the finite resolution 
and finite spatial partition. This practical fact is certainly an important factor  on 
the saturation behavior of {\bf Tts} and the {\bf TTts}.
d) Time series {\bf Ekg}. This STSP is similar to that of time series {\bf Ekg} but with a slower saturation.}}

\end{figure}

\begin{figure}[h]

\vspace{-50pt}

\centerline{\hspace{-3.3mm}
\epsfxsize=8cm\epsfbox{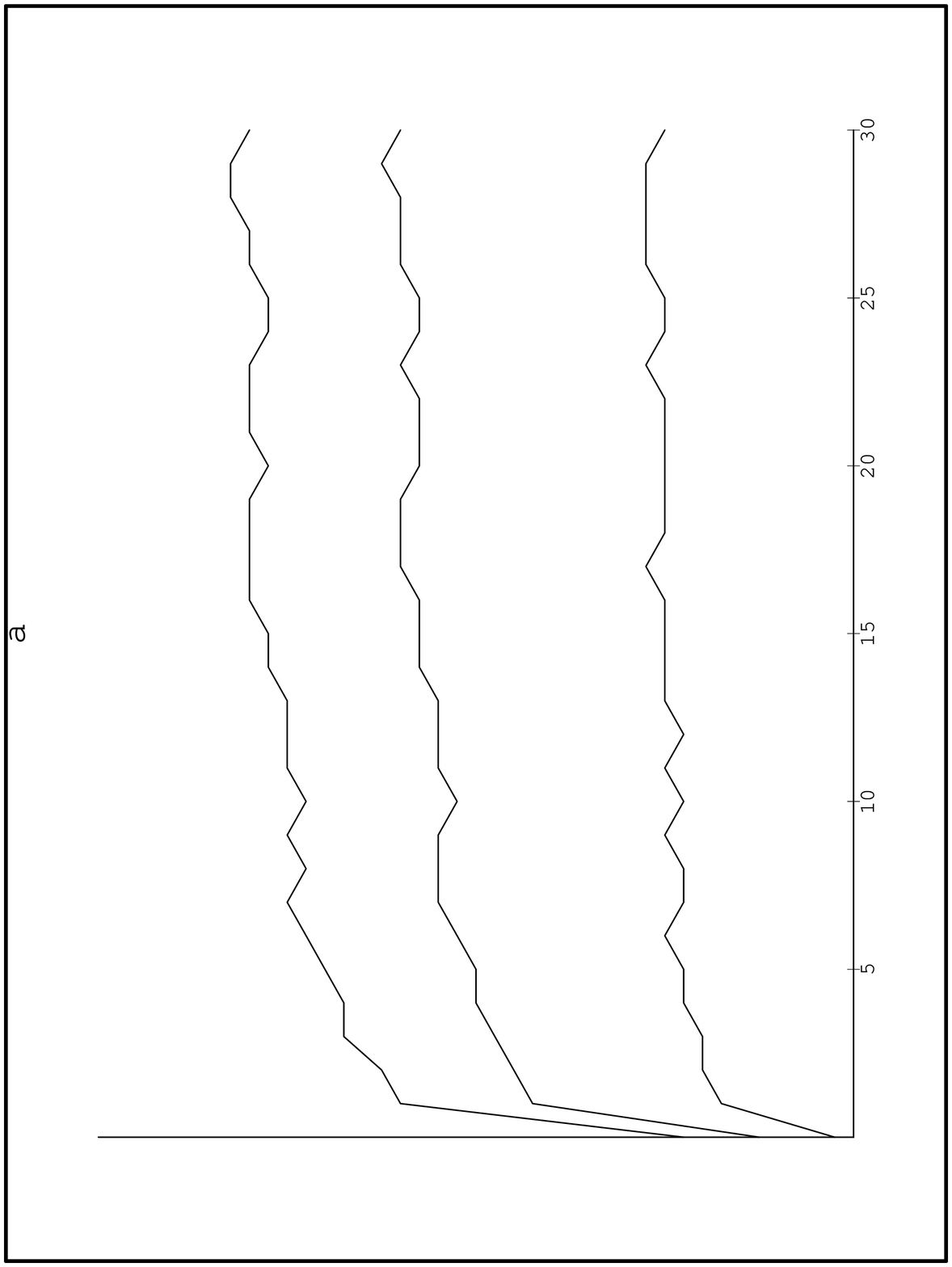}
\hspace{-1cm}
\epsfxsize=8cm\epsfbox{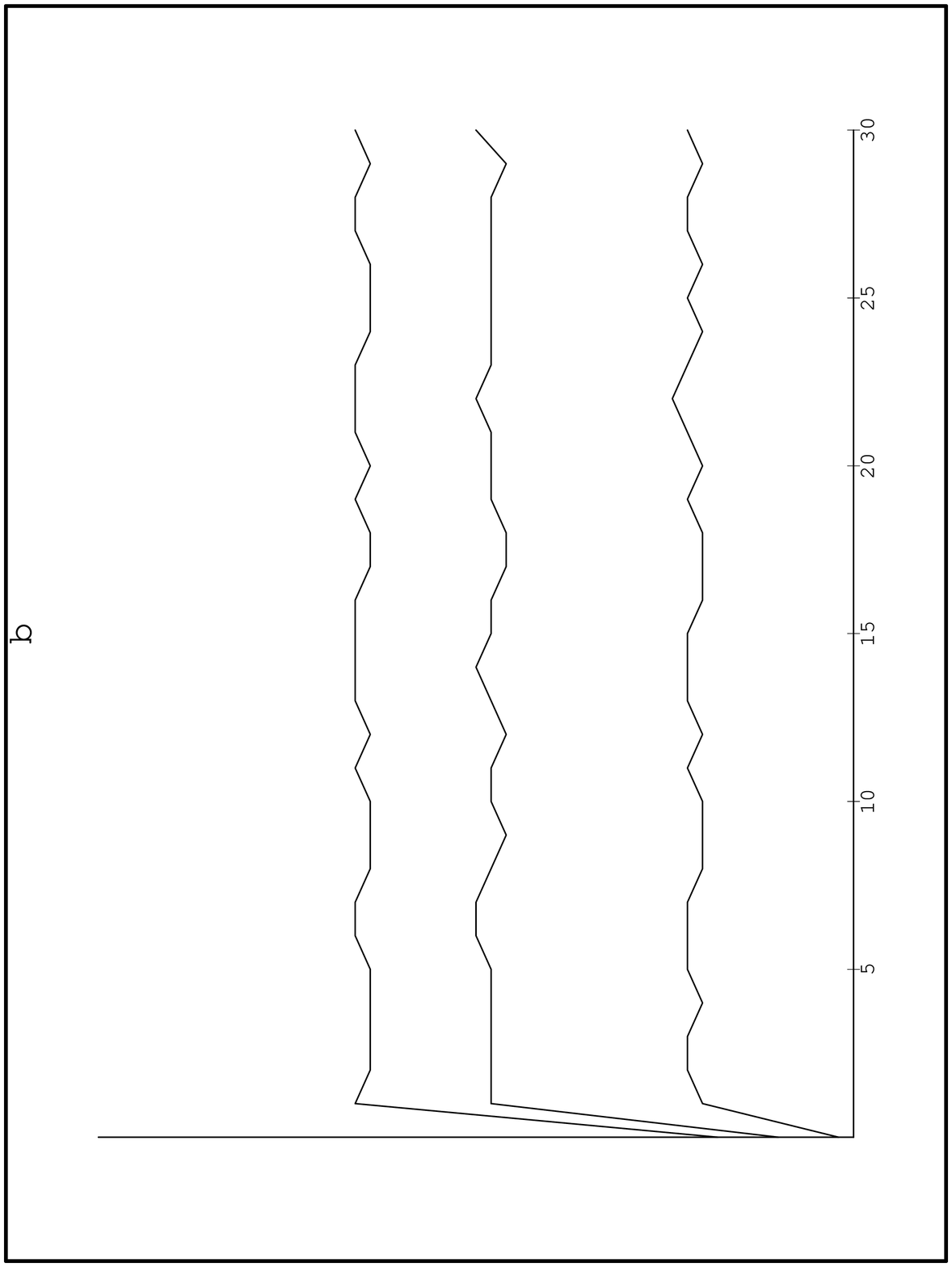}}

\vspace{.0001cm}
\centerline{\hspace{-3.3mm}
\epsfxsize=8cm\epsfbox{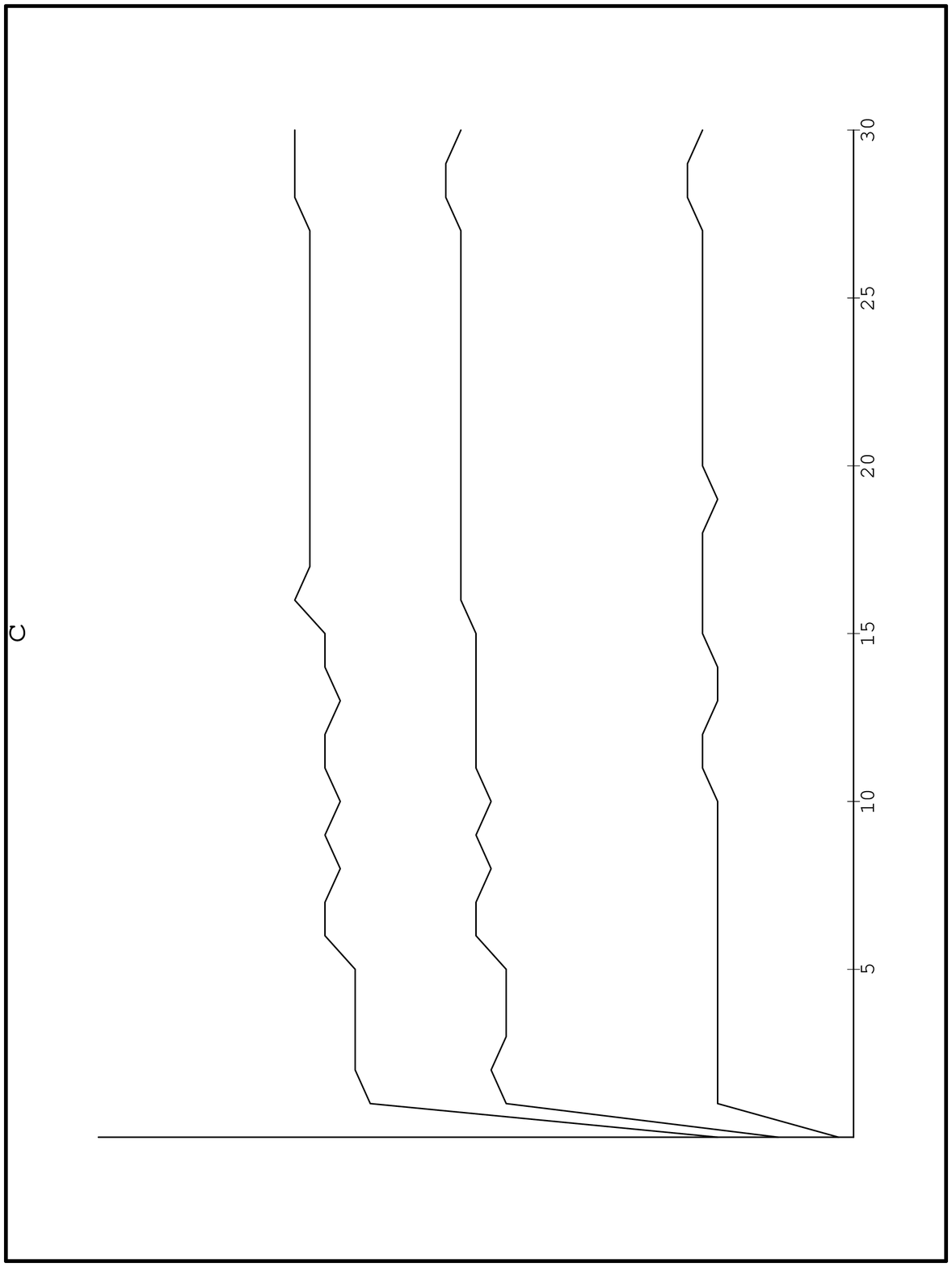}
\hspace{-1cm}
\epsfxsize=8cm\epsfbox{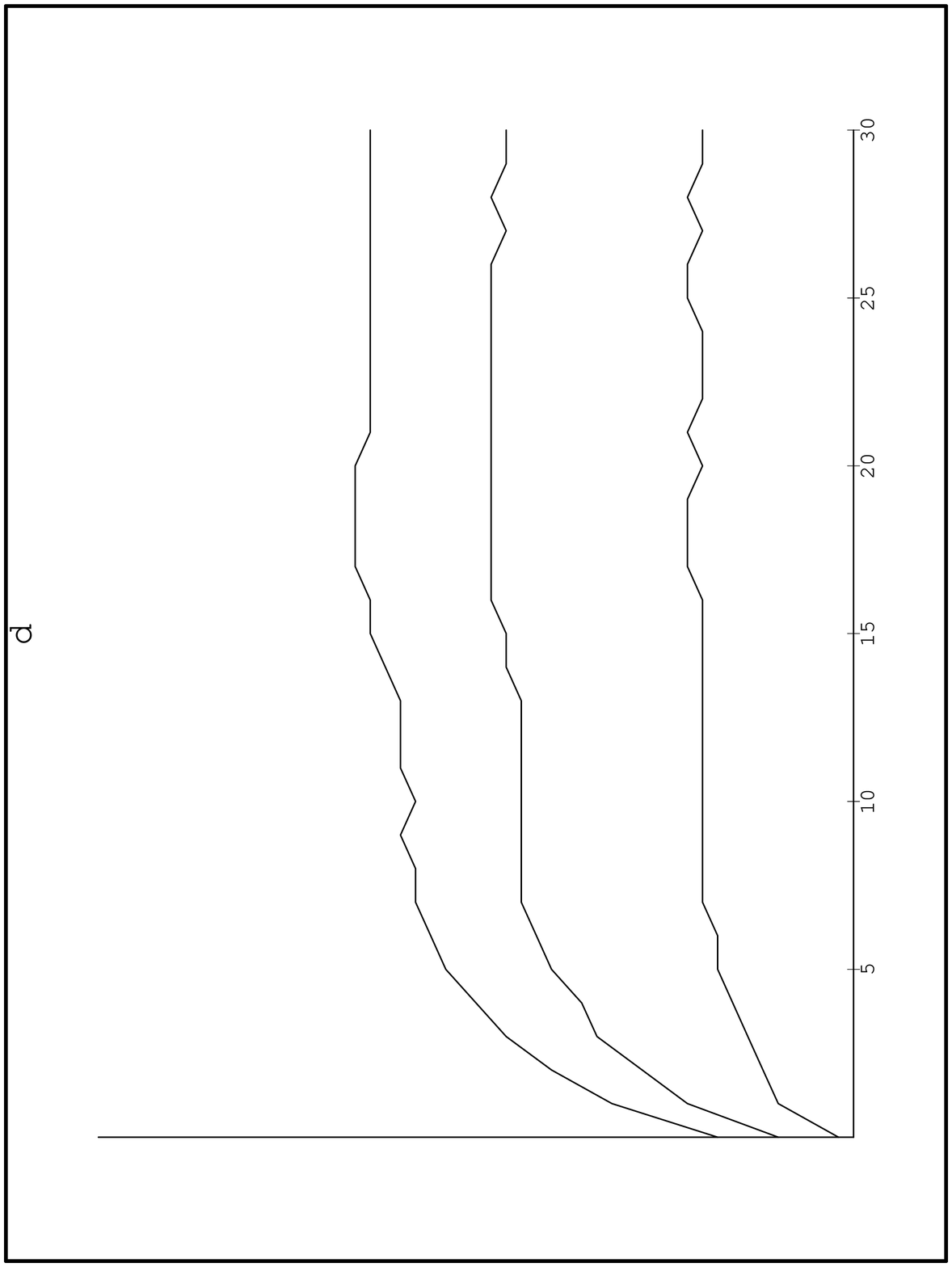}}

\vspace*{-0.3cm}


\caption{{\footnotesize Space Time Separation Plot (STSP), for {\bf Tts} and some representative
 {\bf TTts}. The interpretation of the plot and parameter values are explained in figure 5.
 a) {\bf Tts}. b) {\bf TTts} 5, representative of those {\bf TTts} with
 uncorrelated and stochastic-like plots: 2,3,5,6,7,8,11,12,13 and 18. 
 c) {\bf TTts} 21, representative of those {\bf TTts} with indications of
 determinism and stationarity: 9,19,20,21 and 22. d) 
{\bf TTts} 4, representative of those {\bf TTts} with more clear deterministic and
 stationary results: 4,10,14,15,16 and 17.}}

\end{figure}

Formally speaking, for a strongly stationary time series the results obtained with different 
methods that use some kind of statistics to estimate stationarity, must not depend on the 
embedding parameters. This is because in dynamical systems strong stationarity means that 
in each conceivable embedding space the statistical properties of the phase flows referring 
to different pieces of the time series are the same \cite{WKP98}. However, in practice and 
in cases of less stronger but still valuable stationarity, the estimates of such stationarity 
may depend on the embedding parameters (at least for values beyond certain ranges), 
among others aspects of the particular system and the 
time series at hand.   

\section{ Summary and Discussion}

From inspection of the plot of the data points as a function of time and from all the seven methods applied 
to estimate stationarity, none of the studied time series presented indications of bursts as evidence of 
possible intermittency. Of the seven test time series only {\bf Pnoise}, 
{\bf Brownian}, {\bf Eeg} and {\bf Ekg} presented trends 
measured by a polynomial fit of order two. Indications of nonstationarity given by relatively large $t'$, 
larger than 10 sampling times, are evident for the time series {\bf Pnoise} and {\bf Brownian}. The real world test 
time series {\bf Eeg} and {\bf Ekg} have $t'$ similar to that of the time series {\bf Chaos} which may be representative 
of certain determinism. Indications of nonstationarity represented by dominant low frequencies persistent 
to zero frequency in the power spectrum (PS), are observed for the test time series {\bf Brownian} and less clear 
for the time series {\bf Pnoise}, {\bf Eeg} and {\bf Ekg}. The dominant low frequency PS of the time series {\bf Chaos} is not 
persistent to frequency zero corresponding to a stationary-like nondivergent PS. The time series {\bf Noise} 
and {\bf Random}  have the expected flat PS with no dominant frequency and present the stochastic characteristic 
$t'\sim 1$ and $C(t)\sim 0$ for all $t>1$. Qualitatively, PS for the time series {\bf Eeg} and {\bf Ekg} 
seems to be between 
the power law behavior of the time series {\bf Pnoise} and the exponential behavior of the time series 
{\bf Chaos} with a nonzero dominant frequency.  
{\bf Tts} present clear evidences of nonstationarity with a marked trend and a large correlation time $t'\sim 76$ 
sampling times. The power spectrum of the {\bf Tts} is very similar to that of the time series {\bf Pnoise} and 
persistent to zero frequency as another strong indication of nonstationarity. However, the periodogram 
of the {\bf Tts} differs importantly from that of the time series {\bf Pnoise} although, in the direction of the 
time series {\bf Brownian}. We recall that these two correlated stochastic time series are very common in nature 
and can dominate the dynamical information of {\bf Ots}.

The twenty one {\bf TTts} selected by their improved weak stationarity compared with {\bf Tts}, present interesting 
and varied results. {\bf TTts} 4,9,10, 19, and 20 were not 
explicitly detrended by any method and present the expected similar nonstationary results to {\bf Tts} with 
evident trends and $t'$ larger than 10 sampling times. {\bf TTts} 14, 15, 16, 17, 21 and 22 have  $1<< t' <10$ 
indicating better stationarity and deterministic-like correlations. The remaining ten {\bf TTts} have $t'\sim 1$ 
and $C(t)\sim 0$ for all $t>1$ characteristic of uncorrelated stochastic time series. The periodograms of the 
{\bf TTts} 2, 3, 5, 6, 7, 8, 11, 12 and 13 are all along the 45-degree diagonal also characteristic of 
uncorrelated stochastic time series. From the remaining {\bf TTts} only {\bf TTts} 4, 10, 19 and 20 present dominant 
low frequencies persistent to zero frequency which is an indication of nonstationarity. Therefore, 
the {\bf TTts} 14, 15, 16, 17, 18, 21 and 22 seem to be deterministic and stationary from a linear point of 
view given by PS and $C(t)$.

The results of the nonlinear and complexity measures of such small test time series verify but do not 
demonstrate the corresponding stationary and determinism or stochastic nature of the five test time series 
{\bf Chaos}, {\bf Noise}, {\bf Pnoise}, {\bf Brownian} and {\bf Random} . The real world test time series 
{\bf Eeg} and {\bf Ekg} present indications of determinism and in the first case certain stationarity
depending the method used. In the qualitative framework defined by these results of the nonlinear statistics for the test 
time series, {\bf Tts} presents indications of determinism, or at least indications of correlated nonstationary 
time series with important differences between the first and second half. Notably, the first half of {\bf Tts} is more 
uncorrelated and the second half is less stationary. The two groups {\bf TTts} 2, 3, 5, 6, 7, 8, 11, 12, 13 and 18, 
and {\bf TTts} 9, 19, 20, 21 and 22, present uncorrelated stochastic-like results which are less evident for the 
second group. The remaining {\bf TTts} seem more deterministic but their indications of stationarity vary and 
depend on the nonlinear statistic used. No meaningful conclusions of stationarity can be made from the 
results of nonlinear statistics.

The nonstationarity originated by nonautonomy or by characteristic times of the system of the order of the 
time span of the time series specially identified by RDP, are not observed for the time series {\bf Chaos} and 
the expected overall constant RDP is obtained. Apart from the real world time series {\bf Eeg} and {\bf Ekg}, the 
other four test time series also present the expected results. The stationarity of these 
two real world time series has been improved by choosing a short enough time span of the time series to reduce 
nonstationary effects of larger time variations. For {\bf Tts} we do not have the possibility of adjusting the time 
span without reducing the already small number of data points. Therefore, the clear nonstationarity observed 
in its monotonically decreasing RDP cannot be avoided nor even reduced in this way, the time series has to 
be treated. The apparently stationarity observed in the RDP of the {\bf TTts} 2, 3, 5, 6, 7, 8, 11, 12, 13 and 18 
corresponds to the uncorrelated stochastic-like time series to which {\bf Tts} was reduced to by the corresponding 
treatments. In the case of {\bf TTts} 4, 9, 10, 19 and 20 the nonstationarity of {\bf Tts} has not been improved from 
the point of view of the RDP. The remaining {\bf TTts} 14, 15, 16, 17, 21 and 22 present improved stationarity 
without becoming uncorrelated stochastic-like time series.

The comparison of the results of CPE for the seven test time series with those of the {\bf Tts}, permits us to interpret 
the CPE corresponding to {\bf Tts} as a possibly deterministic but nonstationary time series in its second half. The 
first half has strong indications of weakly correlated stochastic time series as a possible consequence of 
dominating contaminations. The {\bf TTts} with better stationary deterministic indications from the point of view 
of the CPE are 14, 15, 18, 20, 21 and 22. In all of these cases the nonstationarity of the {\bf Tts} has been reduced in 
50\% to 80\% by comparing values of CPE.

From the point of view of the STSP and with the exception of the time series {\bf Brownian} which does not saturate 
at any time, the remaining six test time series can all be considered stationary when temporal correlations 
smaller than 20 sampling times are ruled out. Very small saturation times are 
characteristic of uncorrelated stochastic 
time series. STSP identify short term correlations but it does not easily identify long term correlations 
mostly generated by oversampling and certain contaminations. Thus, the long sampling time and low resolution 
of the {\bf Tts} may be some of the reasons why the corresponding STSP does not saturate for longer times. The 
results of STSP for all {\bf TTts} are therefore very similar to those of the {\bf Tts}. 

\section{ Conclusions}

The limited capacity of obtaining good estimates of stationarity with the different methods when 
applied to {\bf Tts}, 
has been qualitatively improved by the framework of reference given by the test time series and the {\bf TTts}.
The characteristic limitations imposed by {\bf Ots} and in particular by {\bf Tts}, do not permit to obtain 
quantitative and precise meaningful results from a detailed analysis of such time series with any particular
method. Therefore, the construction of a qualitative criteria with increasing accuracy as more information is 
acquired, may provide a quantitative measure with a defined 
accuracy as is the spirit of the statistical perspective.  
The formal and intuitive arguments relating the problems of contamination and stationarity has been observed 
in the concrete case of {\bf Tts}. The intended goal of obtaining 
a meaningful stationary and deterministic time 
series from an {\bf Ots}, in particular {\bf Tts}, becomes a conflict 
between the two desirable qualities of the treated 
time series. This problem becomes evident when we lose 
deterministic content by applying some treatments to 
filter noise and improve the stationarity of the {\bf Tts}. The {\bf TTts} 2, 3, 5, 6, 7, 8, 11, 12 and 18 are clear 
cases of this situation as indicated by most of the analysis performed. The most successful 
results have been obtained by combinations of polynomial detrending with smoothing, filtering with a 
correlation matrix and in particular with nonlinear local filters. The combination of methods which give the best 
results correspond to {\bf TTts} 21, 22, 15, 14, 17 and 16 with decreasing quality as indicated by most of the methods utilized. From these results we corroborate 
that weak stationarity is not always a clear indication for all the possible origins of nonstationarity
considering the strong limitations imposed by the characteristics of {\bf Ots} and particularly for {\bf Tts}.

On one hand, the conflictive problem of improving stationarity and reducing contamination in a {\bf Ots} is closely related with 
the complexity of the underlying process, its interactions with other processes and the 
variety of time scales involved. It may not be possible to solve this problem appropriately. 
On the other hand, such a conflictive situation 
could strongly depend on the process of observation, in such a case certain treatments 
of the {\bf Ots} under study may 
produce a meaningful deterministic and stationary time series with important dynamical information about 
the underlying dynamics.

The stationary and deterministic content obtained for some {\bf TTts} is not completely clear and definitive 
but, it has been improved
compared with {\bf Tts}. The interpretation of the different {\bf TTts}, and to which extent they are meaningful 
improvements of the dynamical information of the original {\bf Tts}, are not problems that can be solved without 
a deep understanding of the phenomenology of the system involved. The goal 
is to reduce the possible explanations from a point of view which manages to capture certain 
exploitable deterministic
characteristics of the {\bf Ots} under study; the {\bf Tts} is the illustrative case. This perspective may provide new
and complementary information looking toward a progress in the understanding of such complex systems 
and their corresponding {\bf Ots}. The  
use of new and improved already existing methods to measure determinism and stationarity in apparently 
stochastic time series, will eventually uncover characteristic features of a particular {\bf Ots}. This work is an initial effort intended to
contribute in the definition and limitations of such an ambitious goal as a part of many efforts 
currently in progress with different perspectives. This new perspective pretends to create a 
meaningful framework of applicability of important and mature methods to {\bf Ots} which have been extensively
tested in situations where important knowledge of the system and good quality
of the corresponding time series are available. In such cases the correct application and 
interpretation of results of the different methods are fairly unambiguous although, there are a lot
of improvements and new ideas to be developed.    

\vspace{.5cm}
\noindent
{\bf Acknowledgments} I thank Colciencias (Colombia) and the Universidad Antonio Nari\~{n}o of 
Santa Fe de Bogot\'a, Colombia for their collaboration.


\begin{thebibliography}{99}
\bibitem{Nota1}    These results may be an artifact of the reconstruction process and the
 norm used as indicated by the minimum probability $p$, of finding identical 
consecutive reconstructed vectors. For $\tau =1$ the probability $p$ is
 very small for large $m$ however, it is not small for small $m$ and 
decreases slowly for increasing but small $m$, $p\geq (m-1)/m^2$. This work is 
in progress.
\end{thebibliography}
\end{document}